\documentclass{article}
\usepackage{natbib}
\usepackage[affil-it]{authblk}
\usepackage[english]{babel}

\usepackage[letterpaper,top=2cm,bottom=2cm,left=3cm,right=3cm,marginparwidth=1.75cm]{geometry}

\usepackage{amsthm}
\usepackage{graphicx} 
\usepackage{longtable}

\usepackage{booktabs,multirow,tabularx} 
\usepackage{multicol} 
\usepackage{float}
\usepackage{amsmath, amssymb, mathtools}
\usepackage[colorlinks=true, allcolors=blue, breaklinks=true]{hyperref}
\usepackage{makecell}
\usepackage{subcaption}
\usepackage{caption}
\usepackage[figurename=Fig., labelsep=space, labelfont=bf]{caption}
\usepackage{setspace}
\usepackage{tikz}
\usepackage{tikz-qtree}
\setcitestyle{round}
\usepackage{multirow}
\usepackage{xcolor}
\usepackage[affil-it]{authblk}

\newtheorem*{proofoutline}{Proof (outline)}
\newtheorem*{Cor}{Corollary}

\usepackage{algorithm}
\usepackage{algpseudocodex}
\algrenewcommand\algorithmicrequire{\textbf{Input:}}
\algrenewcommand\algorithmicensure{\textbf{Output:}}

\title{A heuristic search algorithm for discovering large \\ Condorcet domains}

\author[1]{Bei Zhou \thanks{Corresponding author: bei.zhou@qmul.ac.uk}}
\author[2]{S\o ren Riis }
\affil[1, 2]{Queen Mary University of London}

\date{}
\begin{document}
\maketitle

\begin{abstract}

The study of large Condorcet domains (CD) has been a significant area of interest in voting theory. In this paper, our goal is to search for large CDs that are hitherto unknown. With a straightforward combinatorial definition, searching for large CDs is naturally suited for algorithmic optimisations. For each value of $n>2$, one can ask for the size of the largest CD, thus finding the largest CDs provides an important benchmark for heuristic-based combinatorial optimisation algorithms. Despite extensive research over the past three decades, the CD sizes identified in 1996 remain the best known for many values of $n$. When $n>8$, conducting an exhaustive search becomes computationally unfeasible, thereby prompting the use of heuristic methods. To address this, we developed a novel heuristic search algorithm in which a specially designed heuristic function, backed by a lookup database, directs the search towards promising branches in the search tree. Our algorithm found new large CDs of size 1082 (surpassing the previous record of 1069) for $n$=10, and 2349 (improving the previous 2324) for $n$=11. Notably, these newly discovered CDs exhibit characteristics distinct from those of known CDs.

\end{abstract}

\section{Introduction}
\label{sec:intro}

In social choice theory, the study of preference aggregation and decision-making often grapples with majority rule. For each value of $n$ (representing the number of candidates or alternatives), a Condorcet Domain (CD) is a collection of linear orders (permutations). Crucially, for a CD, every triple of candidates satisfies a specific never condition. This condition, often represented symbolically as $xNy$, stipulates certain forbidden rankings of candidates. This ensures that cyclical majorities, a situation where no candidate can emerge as a consistent winner in pairwise comparisons, are avoided. 

The investigation of CDs, sets of linear orders with an acyclic pairwise majority relation, has been a subject of interest for mathematicians, economists, and mathematical social scientists since the 1950s \citep{danilov2013maximal,fishburn2002acyclic}. CDs also have applications in the Arrovian aggregation and social choice theory \citep{bruner2014likelihood}. In social choice theory, a Condorcet winner wins the majority of votes against each of the other candidates in a pairwise comparison \citep{monjardet2005social}. However, no such candidate exists, and this is where the CDs come into play.

In the 1970s and 1980s, researchers initiated studies into the potential magnitude of CDs \citep{fishburn1997acyclic}. This exploration led to a significant breakthrough in 1996 when Fishburn discovered many of the largest CDs, a record that has remained unchallenged for nearly three decades. Since then, researchers have continued the quest to find large CDs \citep{danilov2012condorcet, fishburn1997acyclic, galambos2008acyclic, karpov2022constructing, monjardet2009acyclic, leedham2024largest}, related to the goal of optimising personal preference freedom while adhering to the requirement of a transitive majority relationship \citep{raynaud1982individual, puppe2024maximal}. In operations search, it is a complex, unsolved combinatorial optimisation problem. Studying CDs has important implications for voting systems and democratic decision-making processes. By identifying large CDs, researchers can gain insights into the properties of voting systems that are more likely to produce outcomes that reflect the electorate's preferences, the strengths and weaknesses of different voting systems, and democratic decision-making processes \citep{lackner2017likelihood}. Thus, a key effort in the literature is identifying large Condorcet domains with a paramount objective to uncover the largest CDs \citep{puppe2024maximal,karpov2022structured,leedham2024largest}. To this end, \citet{PF:1996} introduced the size function to record the largest CDs on $n$ alternatives
\[
f(n)=\max \{\lvert D \rvert: \text{$D$ is a Condorcet domain on a set of $n$ alternatives}\}.
\]
Additionally, \citet{karpovslinko} introduced the following functions:
\[
h(n)=\max \{\lvert D \rvert: \text{$D$ is a peak-pit Condorcet domain on a set of $n$ alternatives}\}.
\]

One important class of large CDs is based on Fishburn's alternating scheme, which involves alternating between two never rules 2N1 and 2N3 on 3-alternative subsets of candidates (commonly referred as triples) and has been used to construct many of the largest known CDs. The domains built by the alternating schemes are referred to as Fishburn domains \citep{leedham2024largest} (see Sect.~\ref{sec:preliminaries} for details). Subsequent research by \citet{galambos2008acyclic,danilov2012condorcet,monjardet2009acyclic} and  \citet{karpov2022symmetric} extended and refined this work. Despite extensive work, these Fishburn domains and their design play a central role in constructing large CDs. For instance, the approach introduced in \cite{karpov2022constructing} was built upon the Fishburn domains to construct new CDs of record-breaking CDs for $n \geq 13$, but it does not apply for $n < 13$. \citet{karpov2023set} introduced the set-alternating schemes that could construct CDs whose size is larger than the Fishburn domains on $n \geq 16$. Their new large CDs also improved the asymptotic lower bound for the size of the largest Condorcet domains \citep{karpovslinko, karpov2023set}. 

Another line of studies focuses on developing efficient search algorithms to generate CDs. \citet{akello2023condorcet} developed a custom search algorithm that generated all the maximal non-isomorphic CD on seven alternatives and found the largest CD on eight alternatives \citep{leedham2024largest} but it does not scale to a larger number of alternatives. \citet{markstrom2024arrow} designed a depth-first search algorithm to generate non-isomorphic maximal Arrow’s single-peaks domains on 9 alternatives. On a higher number of alternatives, completing their search requires a prohibitively demanding amount of computation. In contrast with these existing search algorithms, our algorithm is universal in the sense that it is applicable to any number of alternatives, and it does not aim to compute a complete set of CDs but focuses on the large ones. 

Given the fact that Fishburn domains are not the largest domain on eight alternative \citep{leedham2024largest} and $n \geq 13$ alternatives \citep{karpovslinko}, it is a valid conjecture that for $9 < n < 13$ they are also not the largest. However, existing search algorithms failed to find such CDs. One of our contributions is on the discovery of many CDs larger than the Fishburn domain on $n=10$ and $11$ alternatives. It is worth noting that on $n \geq 9$ alternatives, finding the largest Condorcet domain remains an open problem.

In this paper, we present a heuristic search algorithm for generating and discovering large CDs on $n \geq 6$ alternatives. We first pinpoint the challenges associated with finding large CDs and demonstrate that these challenges pose significant obstacles for common learning algorithms applicable to tackle complex combinatorial optimization problems to find large CDs, especially when dealing with a substantial number of alternatives. This motivated us to develop a heuristic search algorithm tailored to overcome these challenges. Our search algorithm employs an efficient heuristic function that evaluates the goodness of partial CDs based on the size of restriction on 5 alternatives. This idea is grounded on our empirical finding that many locally large restricted CDs also tend to be large, indicating the existence of a linear relationship between the size of a CD and the size of its restriction. Our heuristic function exploits this property and uses a database that contains a complete set of pre-calculated CD restriction information on five alternatives. This idea resonates with the core concept in dynamic programming \citep{bellman1966dynamic} where a problem is broken into smaller, overlapping subproblems, and instead of solving each subproblem independently, solutions to these subproblems are pre-computed and then stored, so they can be reused when the same subproblem arises. 

We further report on new record-breaking CDs on $n=10$ alternatives (size of 1082) and $n=11$ alternatives (size of 2349). This discovery has further improved the largest CD since 1996 and opens new directions for research in the theory of CDs, with implications for the theoretical study of voting systems.

The largest domains our search algorithm found on the number of alternatives ranging from 6 to 11 are presented in Table~\ref{table1}. It is worth noting that along with the new improved lower bound CDs on 10 and 11 alternatives, it also discovered some CDs whose size is larger than those of Fishburn domains for $n=10$ and $n=11$ that were the largest known CDs on these number of alternatives.

\begin{table}[h]
\centering
\small
\captionsetup{width=.6\textwidth}
\caption{The largest domains discovered by our search algorithm. Our work found new improved lower bound CDs of size 1082 and 2349 on 10 and 11 alternatives, respectively and rediscovered the largest CD of size 224 on eight alternatives. }
\label{table1}
\begin{tabular}{cccc}
\toprule
\textbf{n} & \textbf{Fishburn domain} & \textbf{$f(n)$} & \textbf{$h(n)$} \\ 
\midrule
6 & 45 & 45 & 45\\
7 & 100 & 100 & 100\\
8 & 222 & \underline{224} & \underline{224} \\
9 & 488 & 488 & 488 \\
10 & 1069 & \textbf{1082} & \textbf{1082} \\
11 & 2324 & \textbf{2349} & \textbf{2349} \\
 \bottomrule
\end{tabular}
\end{table}

This paper is organised as follows. In Sect.~\ref{sec:preliminaries} we provide the preliminary and background knowledge about Condorcet domains from the operations research perspectives, along with defining and explaining the notions and terminologies used throughout this paper. Sect.~\ref{sec:suboptimality} illustrates the challenges facing a wide range of existing learning algorithms in finding large CDs and analyses their strengths and weaknesses in addressing this problem. Sect.~\ref{sec:algorithm} introduces a heuristic function and shows its effectiveness in establishing a linear relationship between a CD and restrictions, and further elaborates on integrating it into the heuristic search algorithm. In Sect.~\ref{sec:experiments}, we demonstrate the importance of the ordering of triples by which they are assigned and propose a new triple order that contributes partially to the success of the search algorithm. Its detailed configurations, along with an approach that enables it to run on thousands of CPU cores in parallel, are also discussed here. Sect.~\ref{sec:analysis} contains a detailed analysis of the restrictions of the new large CDs we discovered on 10 and 11 alternatives, where we found Ramsey’s Theorem \citep{ramsey} applies. We conclude our work in Sect.~\ref{sec:conclusion}.

\section{Preliminaries and background}
\label{sec:preliminaries}
Let $X_n=[n]=\{1, \dots, n\}$ be a finite set of numbers, in which each element in the set is referred to as an \emph{alternative} or candidate in voting theory. A linear order is a set of alternatives with an ordering, where the comparability and transitivity hold. Here the comparability property ensures that any pair of alternatives can be compared, and the transitivity property ensures that the ordering of alternatives is consistent.  Let $L(X_n)$ denote the set of all linear orders over $X_n$.  In voting theory, voters have to make a collective choice of which alternative to choose. Each voter can be referred to as an agent. Given $N$ agents, each agent $i \in N$ has a preference order $P_i$ in $X_n$ (each preference order is a linear order and can also be viewed as a \emph{permutation}). For brevity, we represent a preference order as a string, e.g. $12 \dots n$ means $1$ is the most favoured alternative, and $n$ is the least. Note that in this paper, we use preference order, linear order and permutation interchangeably. 

A subset of preference orders $D\subseteq L(X)$ is called a \emph{domain} of preference orders or permutations. A domain $D$ is a \emph{Condorcet domain} if whenever the preferences of all agents belong to the domain, the majority relation of the preference profile with an odd number of agents is transitive. Thus a Condorcet domain is essentially a set of permutations where there is always a clear majority preference for any pair of alternatives.

\cite{Sen1966} proved that for each Condorcet domain, restriction of it to each triple of alternatives satisfies a never condition $iNj$, $i,j\in [3]$. A never condition is also called a never rule or simply a rule. A triple consists of 3 alternatives $(x, y, z)$ where $x<y<z$ and $x, y, z \in [n]$. $iNj$ dictates that $i^{th}$ alternative from the triple does not fill in $j^{th}$ place within this triple in each order from the domain. $i$ takes on the value 1, 2, or 3, representing the smallest, middle, or largest alternatives in the triple, respectively. Similarly, the $j$ can be the value 1, 2, or 3, corresponding to the first, second, or last position in the triple. For instance, triple (3, 4, 5) on five alternatives with rule 1N3 dictates that the smallest alternative, 3, cannot present at the last position, forbidding the permutation [4, 2, 5, 1, 3] from presenting in the resulting CD. There are, in total, nine never rules (1N3, 3N1, 2N1, 2N3, 1N2, 3N2, 1N1, 2N2, 3N3).   

A CD's size is defined by the number of permutations contained within it. For any $n \in \mathbb Z_{\ge 3}$ alternatives, a CD is associated with a set of $m={n \choose 3}$ triples, all of which must be assigned with one of the nine rules. A domain is not a CD if there are unassigned triples from which it is created, and we instead call it a \emph{partial} CD. To avoid repetition, we use the acronym TRS to refer to a list of triples and their assigned rule. A TRS produces a CD if all its triples are assigned with a rule. Otherwise, it is a partial CD.

A domain which satisfies a never condition of the form $xN3$ for every triple is called \emph{Arrow's single-peaked domain} \citep{arrow63}. A domain $D$ is a \emph{peak-pit} domain if, for each triple of alternatives, the restriction of the domain to this triple is either single-peaked (satisfying never rules $iN3$ ) or single-dipped (satisfying never rules $iN1$).
A Condorcet domain $D$ is \emph{unitary} if it contains order $123\ldots n$. Since the focus of most of the studies is on unitary CDs, it is sufficient to consider the six rules 1N3, 3N1, 2N1, 2N3, 1N2, and 3N2. 

Furthermore, using the complete set of maximal unitary CDs on four to seven alternatives from \citet{akello2023condorcet}, we calculated the size of the largest CDs where one of their triples uses one of these six rules and the results are displayed in Table~\ref{tab:rule_set}. It shows that applying the rule 1N2 or 3N2 typically leads to poor results in the sense that none of the triples in the largest CDs satisfy either of these two rules.  Besides, the largest CDs on 8 alternatives whose size is 224 also do not use any of these two rules. Thus, we narrow the available rules in our search down to four, which are 1N3, 3N1, 2N1, and 2N3 in our construction. Unless specified, the CDs we will discuss in the following sections are constructed by these four rules. These CDs are by definition peak-pit domains. 

\begin{table}[h]
\captionsetup{width=.7\textwidth}
\caption{The size of the largest CDs with at least one of its triples satisfying each specified rule, on the number of alternatives ranging from 4 to 7. }\label{tab:rule_set}

\centering
\setlength{\tabcolsep}{9pt}
\begin{tabular}{*{7}{c}}
\toprule
& \multicolumn{6}{c}{\textbf{Rules}} \\
\cmidrule(lr){2-7}
\textbf{n} & 1N3 & 3N1 & 2N3 & 2N1 & 1N2 & 3N2   \\
\midrule
4 &  9 & 9  & 8  & 8 & 8 & 8 \\
5 &  20 & 20 & 20  & 20 & 19 & 19 \\
6 & 45 & 45 & 45 & 45 & 42 & 42 \\
7 & 100 & 100 & 100 & 100 & 97 & 97 \\
\bottomrule
\end{tabular}
\end{table}

Two CDs are isomorphic if the alternatives can be re-enumerated in a way that they consist of the same set of permutations \citep{leedham2024largest}. Given two linear orders $P_1$ and $P_2$, a transformation $\pi$ is the application of a mapping that converts $P1$ to $P2$, denoted as $P_2 = \pi (P_1)$. Given a permutation $P \in D$, a transformation  $\pi_P$ converts it to its identity form, i.e. $123...n$ and we use $\pi_P(D)$ to denote a domain obtained from converting all the permutations in $D$ by the transformation $\pi_P$. 

Any non-empty CD $D$ is isomorphic to a unitary CD because we can always select a permutation $P \in D$ and then enumerate the alternatives according to the transformation $\pi_P$ such that the permutation $P$ becomes the identity permutation. Thus, we do not lose any generality by assuming all the CDs we discussed in this paper are unitary. 

To further illustrate the concept of isomorphism on CDs, here we provide an example. Table~\ref{tab:cd_3} presents 6 CDs on three alternatives. Three alternatives can only construct one triple, and thus, each rule assigned to it corresponds to a CD. Three pairs of them are isomorphic. For instance, $D_1=\{123, 132, 213, 312\}$ and $D_2=\{123, 213, 231, 321\}$ are isomorphic because if we define $\pi_{132}$ that maps 132 to the identify permutation 123 (\textit{i.e} $\pi_{132}(1)=1, \pi_{132}(2)=3, \pi_{132}(3)=2$), then $\pi_{132}({D_1})=D_2$, thus $D_1$ and $D_2$ are isomorphic. 

\begin{table}[H]
\captionsetup{width=.8\textwidth}
\caption{The Condorcet domains for three alternatives. Each rule assigned to the triple (1, 2, 3) is associated with a CD. The resulting CDs fall into three isomorphic classes. }
    \renewcommand{\arraystretch}{1.3}
    \centering
    \small
    \begin{tabular}{cccc}
    \toprule
   \textbf{Triple} & \textbf{Rule assigned} & \textbf{Condorcet domains} & \\

    \midrule
    \multirow{6}{*}{(1, 2, 3)}
    & 1N3 &
    \multirow{2}{*}{
    $\begin{rcases*}
        \begin{tabular}{cccc}
             123 & 132 & 213 & 312   \\
             123 & 213 & 231 & 321 \\
        \end{tabular}  
     \end{rcases*}$ 
    } & \multirow{2}{*}{isomorphic}\\
    & 2N3 & \\

    \cmidrule(l){2-4}

    & 3N1 &
    \multirow{2}{*}{
    $\begin{rcases*}
        \begin{tabular}{cccc}
             123 & 132 & 213 & 231  \\
              123 & 132 & 312 & 321\\
        \end{tabular}  
     \end{rcases*}$ 
    } & \multirow{2}{*}{isomorphic}\\
    & 2N1 & \\

    \cmidrule(l){2-4}

    & 1N2 &
    \multirow{2}{*}{
    $\begin{rcases*}
        \begin{tabular}{cccc}
             123 & 132 & 231 & 321 \\
             123 & 213 & 312 & 321\\
        \end{tabular}  
     \end{rcases*}$ 
    } & \multirow{2}{*}{isomorphic}\\
    & 3N2 & \\
    
    \bottomrule
    
\end{tabular}
\label{tab:cd_3}
\end{table}

The \emph{restriction} of a domain $D$ to a subset $A\subset X$ is a domain with the set of linear orders from $L(A)$ obtained by restricting each linear order from $D$ to $A$ \citep{karpov2024local}.  Given $k\in[3, n-1]$ alternatives, a CD on $n$ alternatives can have restrictions on $k$ alternatives by removing the alternatives in every linear order (permutation) within the CD that are not these $k$ alternatives and retaining only one of the duplicated resulting liner orders. The resulting restrictions contain the linear orders that only contain these $k$ alternatives.

Analogously, given $k\in[3, n-1]$ alternatives, a TRS for $n$ alternatives can also be restricted to $k\in[3, n-1]$ subset triples by retaining the triples that are included in these $k$ alternatives, which is equivalent to removing the triples that include least one of $(n-k)$ alternatives that are not in these $k$ alternatives. The rules are transferred from the original triples to the subset triples. A set of triples for $n$ alternative, when restricted to $k$ alternatives subset, produces ${n \choose k}$ sets of subset triples, each of which corresponds to one combination of $k$ alternatives from $n$ alternatives. For example, a set of 11 alternatives produces 10 sets of subset triples when restricted to 10 alternative subsets and 55 sets of subsets when restricted to 9 alternative subsets. 

Before our results, the largest CDs for 10 and 11 alternatives were Fishburn domains. A domain $D$ is called \emph{Fishburn domain} if it satisfies the alternating scheme \citep{fishburn1996acyclic}: there exists a linear ordering of alternatives $a_1, \ldots, a_n$ such that for all $i, j, k$ with $1\le i<j<k\le n$ the restriction of the domain to the set $\{a_i,a_j,a_k\}$ is single-peaked if $j$ if is even (odd), and it is single-dipped if $j$ is odd (even). The alternating scheme explicitly uses the parity function. When applying the alternating scheme to the list of triples ($x_i, y_i, z_i$), it dictates that if the middle alternative $y_i$ in a triple is even, it is assigned with rule $2N1$, otherwise $2N3$. Despite its simplicity, it turned out that the list of generated rules built large CDs that had held the records for almost three decades until recent studies discovered some new large CDs on eight alternatives \citep{leedham2024largest} and $n\geq13$ alternatives \citep{karpov2022constructing}.
 
CDs on five alternatives are constructed on 10 triples, as demonstrated in Table~\ref{table:triples_rules_5}, where their rules are assigned by applying the alternating scheme. The resulting Fishburn domain on five alternatives has size 20. The permutations inside the CD are presented in Table~\ref{tab:cd_5}. It is the largest CD on five alternatives. 

\begin{table}[H]
\small
\captionsetup{width=.55\textwidth}
\caption{The list of triples for five alternatives and their rules are assigned according to the alternating scheme.}
\captionsetup{width=.8\textwidth}
\centering
\begin{tabular}{lc}
\toprule
\textbf{Triples} & \textbf{Rules}\\
\midrule
(1, 2, 3) & 2N3 \\
(1, 2, 4) & 2N3 \\
(1, 2, 5) & 2N3 \\
(1, 3, 4) & 2N1 \\
(1, 3, 5) & 2N1 \\
\bottomrule
\end{tabular}
\hspace{1em}
\begin{tabular}{lc}
\toprule
\textbf{Triples} & \textbf{Rules}\\
\midrule
(1, 4, 5) & 2N3 \\
(2, 3, 4) & 2N1 \\
(2, 3, 5) & 2N1 \\
(2, 4, 5) & 2N3 \\
(3, 4, 5) & 2N3 \\
\bottomrule
\end{tabular}

\label{table:triples_rules_5}
\end{table}

\begin{table}[h]
\captionsetup{width=.5\textwidth}
\caption{The Fishburn domain on five alternatives. It consists of 20 permutations built from the triples and rules in Table~\ref{table:triples_rules_5}.}
\small

\centering
\begin{tabular}{cccc}
\toprule
\multicolumn{4}{c}{Fishburn domain on five alternatives} \\
\midrule
12453 & 12435 & 12345 & 54321 \\
45321 & 54231 & 45231 & 42531 \\
24531 & 54213 & 45213 & 42513 \\
42153 & 42135 & 24513 & 24153 \\
24135 & 21453 & 21435 & 21345 \\

\bottomrule
\end{tabular}

\label{tab:cd_5}
\end{table}

The Fishburn domain on five alternatives has five restrictions on four alternatives, whose sizes are 9, 9, 8, 9, 9 as demonstrated in Table~\ref{tab:subset_sizes_5} where each restriction has one alternative removed from the original CD. For example, the restriction on four alternatives produced by removing alternative one from these 20 permutations in this Fishburn domain contains nine permutations. 

\begin{table}[H]
    \centering
    \caption{The 4 alternative restrictions from the Fishburn domain on 5 alternatives. }
    \small
    \begin{tabular}{clcccc}
    \toprule
    \textbf{Removed alternative} & \multicolumn{5}{c}{\textbf{Restrictions}}\\

    \midrule
    \multirow{2}{*}{1} & 4253 & 4523 & 5432 & 2345 & 5423 \\
    & 2453 & 4532 & 4235 & 2435 \\

    \midrule
    \multirow{2}{*}{2} & 4531 & 5413 & 4513 & 5431 & 4135 \\ 
    & 1453 & 4153 & 1345 & 1435\\ 
    
    \midrule
    \multirow{2}{*}{3} &4521 & 4215 & 2145 & 2415 & 1245 \\ 
    & 4251 & 5421 & 2451\\ 

 \midrule
    \multirow{2}{*}{4} & 1235 & 2513 & 2153 & 2135 & 2531 \\ 
    & 1253 & 5213 & 5231 & 5321\\

     \midrule
    \multirow{2}{*}{5} & 2143 & 2134 & 1234 & 2431 & 4213 \\
    & 4231 & 2413 & 1243 & 4321 \\

    \bottomrule
\end{tabular}
\label{tab:subset_sizes_5}
\end{table}

\begin{figure}[H]
    \centering
    \begin{tikzpicture}[scale=1.5]
    
    \centering
    \node[] (0) at (0,0) {\textbf{20}};
    
    \node[] (1) at (-2,-1) {\textbf{9}} ;
    \node[] (2) at (-1,-1) {\textbf{9}};
    \node[] (3) at (0,-1) {\textbf{8}};
    \node[] (4) at (1,-1) {\textbf{9}};
    \node[] (5) at (2,-1) {\textbf{9}};

    \draw[->] (0.south) -- (1.north) node[near end, above]{1};
    \draw[->] (0.south) -- (2.north) node[near end, right]{2};
    \draw[->] (0.south) -- (3.north) node[near end, left]{3};
    \draw[->] (0.south) -- (4.north) node[near end, left]{4};
    \draw[->] (0.south) -- (5.north) node[near end, above]{5};

    \end{tikzpicture}
    \captionsetup{width=.6\textwidth}
    \caption{The restriction sizes for the Fishburn domain on five alternatives when restricted to 4 alternatives.}
    \label{fig:n_5_20_subsets}
\end{figure}
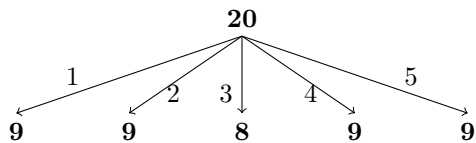

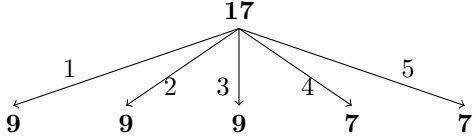
\begin{figure}[H]
    \centering
    \begin{tikzpicture}[scale=1.5]
    
    \centering
    \node[] (0) at (0,0) {\textbf{17}};
    
    \node[] (1) at (-2,-1) {\textbf{9}} ;
    \node[] (2) at (-1,-1) {\textbf{9}};
    \node[] (3) at (0,-1) {\textbf{9}};
    \node[] (4) at (1,-1) {\textbf{7}};
    \node[] (5) at (2,-1) {\textbf{7}};

    \draw[->] (0.south) -- (1.north) node[near end, above]{1};
    \draw[->] (0.south) -- (2.north) node[near end, right]{2};
    \draw[->] (0.south) -- (3.north) node[near end, left]{3};
    \draw[->] (0.south) -- (4.north) node[near end, left]{4};
    \draw[->] (0.south) -- (5.north) node[near end, above]{5};
  
    \end{tikzpicture}
    \captionsetup{width=.6\textwidth}
    \caption{The size of 5 restrictions on 4 alternatives from a five alternative CD of size 17.}
    \label{fig:n_5_17_subsets}
\end{figure}

There is a positive correlation between the size of a CD and the size of its restrictions \citep{karpov2024local}, and to illustrate we use the lattice structure to show the size of a Fishburn domain on five alternatives and the sizes of their restrictions when restricted to 4 alternatives in Fig.~\ref{fig:n_5_20_subsets}. The number sitting aside the arrows is an alternative and indicates that the restriction is built by eliminating that alternative from permutations in the original CD. The largest CD on four alternatives is of size 9. Four restrictions are of the maximum size for four alternatives, and only one CD is of size 8. To make a comparison, we also show in Fig.~\ref{fig:n_5_17_subsets} a CD on five alternatives of size 17 and its restriction sizes on 4 alternatives. Two of the restrictions are of size 7. It is evident that, in this case, a larger CD leads to a set of larger restrictions. We further discovered that this holds true for many CDs in general. Based on this finding, we constructed a heuristic function that captures this correlation. We show in Sect.~\ref{sec:algorithm} that this heuristic function establishes a nearly linear relationship between the size of a CD and a value calculated from an aggregation of its five alternative restriction sizes. 

\section{Suboptimality of existing learning algorithms}
\label{sec:suboptimality}

The search for large CDs is a challenging discrete combinatorial optimisation problem that requires significant effort and faces several challenges where many common learning algorithms struggle. While learning algorithms like reinforcement learning algorithms, genetic algorithms, and local search algorithms have proven their effectiveness in solving some combinatorial optimisation problems \citep{fawzi2022discovering,mankowitz2023faster,radhakrishnan2021evolutionary,el2018particle,lauri2023learning}, they encounter immense difficulty in finding large CDs for $n\geq8$ alternatives. We tested a range of algorithms that can solve discrete combinatorial optimisation problems on finding large CDs. All the programs we used are publicly available\footnote{GitHub. \url{https://github.com/sagebei/cdl/tree/main/algorithms/benckmarking}} and their execution was terminated when no improvement was seen after a reasonable amount of time. The large CDs they found on four to eight alternatives are presented in Table~\ref{tab:cd_benchmark}. Apart from the simple hill climbing algorithm, all the tested learning algorithms found the largest CD on four and five alternatives. But on seven and eight alternatives, none of them found the largest known CDs. This indicates that while these algorithms can find large CDs, there are challenges for them in finding higher lower bounds for CDs on a higher number of alternative alternatives. We summarize these challenges as follows. 

The most prominent difficulty facing them is the exponential growth of the search space. As the number of alternatives increases, the number of potential solutions expands exponentially, making exhaustive exploration impractical or computationally infeasible for $n\geq6$ \citep{leedham2024largest}.
To overcome this issue, prior works have designed search algorithms capable of significantly reducing the search space. \citet{akello2023condorcet} released complete sets of isomorphic CDs on $4 \leq n \leq 7$ alternatives, which were obtained from a search algorithm they developed that produces all the maximum CDs on a given number of alternatives whose size is larger than or equal to a cutoff value. Using a paralleled version of this algorithm on a supercomputer, \citet{leedham2024largest} found the largest CDs on eight alternatives. However, their algorithms are not practically applicable to finding the maximal CD for $n\geq9$ as the search tree that the algorithm needs to build is too large. For a full peak-pit CD of 9 alternatives, there are 84 triples, and thus, when each of them is assigned a rule chosen from a list of 4 never rules, it results in a search space of size $4^{84}$. In practice, though their algorithm can reduce the search space, it is still impossible to search through the reduced search space exhaustively. Additionally, although many structures in the search space could be exploited \citep{puppe2024maximal}, greedy strategies that reduce the search space by narrowing the search to focus solely on large partial CDs failed miserably.  

Consequently, these algorithms often use approximate methods to navigate the vast solution space efficiently. However, the intricate interdependencies among these triples exacerbate the problem. The presence of complex relationships significantly restricts the feasible search space, making it challenging for algorithms to explore all potential solutions effectively. A mistake at any step is likely to spoil the whole solution. 

The existence of many local optima corresponding to suboptimal solutions poses another significant hurdle for these algorithms. Local search algorithms and evolutionary techniques, including genetic algorithms, are particularly prone to getting trapped in local maxima due to their reliance on iterative refinement processes. Escaping these local optima and effectively exploring the entire solution space to converge to the global optimum is a challenging task. From a general understanding of the complete set of CDs on seven alternatives \citep{akello2023condorcet}, many large domains have sizes that are close to the maximum possible, but they represent different local optima. Thus, many learning algorithms get trapped in the swamp of the large domains, making it hard to escape and find the largest domain. 

Calculating the size of the CD on the large number of alternatives is computationally heavy and time-consuming \citep{zhou2023cdl}, leading to another major issue where it is computationally demanding to compute the resulting CD sizes millions of times needed for the learning algorithms to improve on the feedback. 


\begin{table}[H]

\caption{The largest CDs found by a range of learning algorithms. Common learning algorithms are vulnerable to local traps and victims of huge search space in finding large CDs. (DQN: Deep Q learning; PPO: Proximal Policy Optimisation; A2C: Advantage Actor-Critic; EC: Evolutionary Computation; GA: Genetic Algorithms; ES: Evolutionary Strategies; PSO: Particle Swarm Optimisation; HC: Hill Climbing; SA: Simulated Annealing).}
\centering
\small
\begin{tabular}{cccccccccc}
\toprule
 & & \multicolumn{3}{c}{\textbf{Deep RL}} & \multicolumn{3}{c}{\textbf{EC}} & \multicolumn{2}{c}{\textbf{Local Search}} \\
\cmidrule(lr){3-5}\cmidrule(lr){6-8}\cmidrule(lr){9-10}
\textbf{n} & \textbf{Largest size} & \textbf{DQN} & \textbf{PPO} & \textbf{A2C} & \textbf{GA} & \textbf{ES} & \textbf{PSO}  & \textbf{HC} & \textbf{SA} \\
\midrule
4 & 9 & 9 & 9 & 9 & 9 & 9 & 9 & 9 & 9 \\
5 & 20 & 20 & 20 & 20 & 20 & 20 & 20 & 16 & 20 \\
6 & 45 & 44 & 45 & 41 & 44 & 41 & 45 & 41 & 45 \\
7 & 100 & 86 & 94 & 85 & 96 & 81 & 92 & 71 & 97 \\
8 & 224 & 180 & 199 & 181 & 194 & 162 & 182 & 145 & 216 \\
\bottomrule
\end{tabular}

\label{tab:cd_benchmark}
\end{table}

Finding large CDs can be naturally cast as a reinforcement learning problem where it can be modelled as a single-player game. The game starts with an initial position, a list of empty triples with no rule assigned. In each game step, the player assigns a rule to the next triple in the sequence. It ends when all the triples are assigned. There are no intermediate rewards, and the final reward is in proportion to the size of the constructed Condorcet game at the end of the game. In this setting, an RL agent sequentially assigns a rule to the list of ordered triples and gets a reward. 

Deep reinforcement learning (DRL) algorithms, while powerful, still have limitations when it comes to solving hard combinatorial optimisation problems \citep{mazyavkina2021reinforcement, barrett2020exploratory}. 
Deep Q learning algorithms \citep{mnih2015human}, Proximal Policy Optimisation algorithms \citep{schulman2017proximal} and Advantage Actor-Critic algorithms \citep{mnih2016asynchronous} in their original form are prone to converge to local optima as \citet{barrett2020exploratory} described. In our experiments, they all failed at finding the largest CDs on $n\geq7$ alternatives.

Given these identified challenges, we analyze the following factors contributing to this issue. One of the main limitations is the curse of dimensionality, which refers to the exponential increase in the number of possible solutions as the problem size grows, which is the exact problem facing finding the largest CDs. This can make it difficult for DRL algorithms to search through all possible solutions and find the optimal one in a reasonable amount of time. Another limitation is the presence of local optima, which traps the algorithm in a sub-optimal solution. This is particularly true in our case where the reward function is sparse (\textit{i.e.} a reward is given only at the end of the game), making it difficult to differentiate between good and bad solutions. While the size of the partial CDs is a good choice for intermediate rewards, they do not strongly correlate with the size of the resulting full CD and sometimes can be misleading. DRL algorithms also require a lot of computational power, memory, and data to train effectively, which can be a challenge for working with large alternative CDs. Additionally, finding the set of rules that constructs a large CD requires a high degree of precision, which is difficult to achieve with reinforcement learning algorithms. Despite these limitations, DRL algorithms have shown promising results in various applications, and researchers are working to overcome these challenges to improve their performance on hard combinatorial optimisation problems \citep{mazyavkina2021reinforcement,bello2016neural}. 

The AlphaZero-style algorithms, one of the deep reinforcement learning algorithms, has achieved remarkable success on board games like Chess and Go by utilizing a value function\footnote{Note that AlphaZero algorithms also use policy function to cut off poor branches, but it is out of the scope of this paper to discuss the details} that evaluates the current board position without waiting for the game's completion, as described by \cite{silver2018general}. 
The AlphaZero algorithm and its variations do not only apply to board games but are also applicable to solving combinatorial optimisation problems \citep{laterre2018ranked}. For example, \citet{fawzi2022discovering} applied a variation of AlphaZero called AlphaTensor to improve the best-known matrix multiplication algorithms, and \citet{mankowitz2023faster} developed AlphaDev that found fast sorting algorithms, further showcasing the algorithm's potential on tackling hard combinatorial optimisation problems. 

Conceptually, the value function, commonly denoted as $V_{\pi}(s)$ where $s$ is the state the agent is visiting, specifies how good it is in that state when following the policy $\pi$. During the tree search, the search space can be reduced by discarding the branches regarded as unfavourable by the value function. 

However, training or building an accurate optimal value function is challenging. A recent study has shown that when the estimations of the value function are not accurate, \textit{i.e.}, it fails to predict the precise optimal value of states, the direction of the search could be misled to the points where the algorithm fails to learn to play the child game of Nim, one of the classic impartial games \citep{zhou2022impartial}. That and some of the literature it referred to also show that modelling parity function by neural networks on long bitstrings is a non-trivial task. To some extent, the parity issue that characterizes the challenge of learning Nim-like games also arises in CDs. Their study also highlights the difficulty of modelling the parity function using neural networks on long bitstrings, which characterizes the challenges of learning Nim-like games and CDs. As discussed in Sect.~\ref{sec:preliminaries} all previously known CDs have utilized the alternating scheme, which explicitly employs the parity function. 

Overcoming this fundamental difficulty for neural networks is crucial to improve their performance in such domains. While the potential of reinforcement learning algorithms like AlphaZero is promising, the challenges of training accurate value functions must be carefully considered. Our search algorithm circumvents these challenges by applying a handcrafted heuristic (value) function. 

Evolutionary algorithms, including genetic algorithms (GA), evolutionary strategies (ES), and particle swarm optimisations (PSO), though they have achieved notable success in solving combinatorial optimisation problem \citep{radhakrishnan2021evolutionary,oliveto2007time,papadrakakis1998structural,calegari1999taxonomy,casas2022heuristic}, also face specific challenges when applied to finding large CDs. While genetic algorithms (GAs) are a popular optimisation technique, they too have limitations when it comes to solving hard combinatorial optimisation problems \citep{khuri1994evolutionary}. GAs are prone to converge prematurely, meaning the search may stop before the optimal solution is found. This can happen when the search space is too large and the population is not diverse enough \citep{elsayed2014new}. GAs also require several parameters to be set, such as the population size, crossover probability, and mutation rate. Finding the right set of parameters can be difficult, and small changes in these parameters can significantly affect the quality of the solutions. GAs can struggle with scalability when the problem size grows too large. As the search space expands, the time required to evaluate each solution increases, and the population size required to obtain diverse solutions increases exponentially. 

Evolutionary strategy algorithms are often used to solve continuous optimisation problems. Still, they can also be applied to solve discrete combinatorial optimisation problems by optimizing the weights of a neural network that generates the solutions \citep{salimans2017evolution,papadrakakis1998structural}. To this end, they are strong alternatives to reinforcement learning algorithms \citep{salimans2017evolution,majid2023deep}. Particle swarm optimisation algorithms are also designed to solve optimisation problems whose solution space is continuous \citep{el2018particle,tchomte2009particle,selvi2010comparative}, but this limitation can be overcome by using a neural network. 

Local search algorithms are also good candidates for solving hard combinatorial optimisation problems \citep{johnson1988easy}. One of the typical local search algorithms is Simple Hill Climbing (SHL), where the current solution is iteratively modified by making small changes. If the modified solution is better than the current solution, it is accepted as the new current solution. This process continues until no further improvement can be made. It is not hard to see that it is susceptible to getting stuck in local optima. Random Restart Hill Climbing overcomes the limitation by performing multiple hill-climbing searches from different random starting points. Each search is run until it reaches a local optimum, and the best solution among all searches is selected as the final result. 

Simulated annealing is a variant of hill climbing that allows for occasional "downhill" moves to avoid getting trapped in local optima \citep{kirkpatrick1983optimization}. It uses a temperature parameter that controls the probability of accepting worse solutions at the beginning of the search but reduces this probability as the search progresses. This allows the algorithm to explore a wider search space before converging to the optimal solution. Simulated annealing can be computationally expensive, and the time required to evaluate each solution increases as the problem size grows. As a result, SA can struggle with scalability when the problem size becomes too large. 

We experimented with these two local search algorithms, random-restart Hill Climbing and Simulated Annealing with Restarts. As expected, they struggled. On four, five, and six alternatives, the latter found the maximal size of 9, 20, and 45 (both when searching with the four pit-peak rules and all six rules). For seven alternatives, the largest CD size it found was 97, and for eight alternatives, it found a CD of size 216. A major issue is that it is computationally demanding to compute CD sizes the millions of times needed for the algorithm to get enough feedback to efficiently explore the search space and escape from the local trap to reach the global optima. 

The fact that the GA and SA found relatively larger CD for seven and eight alternatives in comparison with other algorithms shows that focusing on a small population of good solutions and further exploring them is a good strategy, and on the contrary, striving to build a good solution from scratch tends to end up in local traps. The utility of value functions in RL algorithms in solving mathematical problems \citep{fawzi2022discovering} and combinatorial optimisation problems \citep{mazyavkina2021reinforcement}, along with the benefit of focusing on a set of good candidate solutions as in GA and SA inspired our heuristic search method in which we designed a handcrafted heuristic function to calculate a value estimating the goodness of a set of partially assigned triples, and the search further explores the ones that have high evaluations.  

\section{The algorithm}
\label{sec:algorithm}

Using the partial CD size as a measure to evaluate it leads to two issues. Calculating the size of a partial CD on a large number of alternatives entails prohibitively costly computation. More importantly, there is no guarantee that a large partial CD could lead to a large full CD, which is why the greedy algorithms stumble, indicating that using the size of a partial CD as a measurement to rank them is erroneous. This promotes the need to develop heuristic functions to evaluate them. A best heuristic function establishes a strong linear relationship between the size of a CD and its value for all the CDs where for any two CDs, $A$ and $B$ if the size of $A$ is larger than that of $B$, the value of $A$ is also higher than that of $B$. However, training or building a best heuristic function is challenging. In this section, we present a search algorithm that applies a manually designed heuristic function using a database that could capture the positive relationship.

\subsection{Database construction}

The fact the size of a CD and the sizes of its restrictions are linearly correlated suggests that the sizes of the restrictions of two CDs can be utilised to measure their relative size. To extrapolate, the largest full CD size reachable from a partial CD is also correlated with the sizes of the largest full restrictions reachable from its restrictions. On a small number of alternatives, we designed an efficient algorithm to find out the size of the largest CD any partial CD can end up with, and store them in a database to enable fast look-up. 

The database contains a set of precomputed (key$\colon$value) pairs where the keys are the states of fully and partially assigned triples for all combinations of rule assignment on five alternatives and their corresponding value is the largest possible full CD size obtainable from a partial CD or the actual CD size for a full CD. The TRS state conception was proposed in \citet{zhou2023cdl}, which refers to the string representation of the rules assigned to a list of triples in an order.  This database enables swiftly obtaining the largest possible size of any five alternative CDs given their state by simply looking it up. An efficient algorithm that constructs the database on $3 \leq n \leq 5$ is given in Algorithm~\ref{alg:database}. Considering the exponentially growing computation complexity of building the database as $n$ grows larger and the memory needed to store the dataset, it is impractical to construct it on $n \geq 6$ alternatives. 

\begin{algorithm}[H]
\caption{Construct the database}\label{alg:database}
\begin{algorithmic}
    \Require{The number of alternatives $n$ where $n \geq 3$ \\
             A set of rules $R$: 2N1, 2N3, 1N3, 3N1 by default} 
    \Ensure{A key-value database $\mathcal{D}$ on $n$ alternatives} 

    \State Add $|R|^{m}$ pairs of (state of full TRS with $R$ $\colon$ their resulting CD size) to $\mathcal{D}$
    
    \For {$i=1, 2, \ldots, m-1$}
        \State $j$ $\leftarrow$ $m-i$ 
        \State $k$ $\leftarrow$ $j \choose m$ 
        \State $\mathbf{R}_{j}$ $\leftarrow$ Cartesian product of $\underbrace{R \times R \times \ldots \times R}_{j}$
        \State $\mathbf{TRS}$ $\leftarrow$ a list of $k$ TRSs each with $j$ triples assigned with 3N1 \Comment{This rule can be any.}
        \For {each TRS in $\mathbf{TRS}$}
            \State $T_{i} \leftarrow$ unassigned triples in TRS
            \State $T_{j} \leftarrow$ assigned triples in TRS
            \For {each $R_j$ in $\mathbf{R}_{j}$}
                \State Reassign $R_{j}$ to $T_{j}$ for TRS
                \State largest\_size $\leftarrow$ 0
                \For {each unassigned triple $t$ in $T_{i}$}
                        \For {each rule $r$ in $R$}
                            \State Assign $r$ to $t$ for TRS
                            \State size $\leftarrow$ look up TRS state in $\mathcal{D}$
                            \If{$\text{size} > \text{largest}\_\text{size}$}:
                                \State largest\_size $\leftarrow$ \text{size}
                            \EndIf
                            \State Assign empty rule to $t$ for TRS \Comment{leave $t$ unassigned}
                        \EndFor
                \EndFor
            \State Add (TRS state $\colon$ largest\_size) pair to $\mathcal{D}$
            \EndFor
        \EndFor
    \EndFor
\end{algorithmic}
\end{algorithm}

\subsection{The heuristic function}

The main objective of a heuristic function is evaluating partial CDs. For a partial CD, the heuristic function should calculate a value in proportion to the largest possible full CD size obtainable from it. That is, for any given CD, the heuristic function should establish a trend in which the larger the full CD that a partial CD can be, the larger its value.  Given that all the CDs for $n \leq 7$ have been studied, we aim to develop the heuristic function that applies to the CDs on $n \geq 8$ alternatives. 

For any TRS on $n$ alternatives where $n\geq6$, the heuristic function given in Equation~(\ref{eq:hf}) takes as input a set complete set of $s( = {n \choose 5})$ subset states and look them up in the database to fetch their values $\mathbf{v}=[v_1, v_2, \ldots, v_s]$ where $v_i\in [1, 20]$. They are then aggregated to obtain a single value estimating the goodness of this TRS. The aggregation is a linear weighted combination of the number of occurrences of the size of each of the five alternative CDs.

\begin{equation}
\label{eq:hf}
\begin{aligned}
f(\mathbf{w}, \mathbf{v}) = \begin{cases}
-1 & \text{if } \min(\mathbf{v}) \leq 15 \\
 \sum\limits_{i=16}^{20} m_{i} \cdot w_{i} & \text{otherwise}
\end{cases} \\
\end{aligned}
\end{equation}

where $m_i$ denotes the number of occurrences for five alternative CDs of size $i$, and $w_i$ is its corresponding weight. To lead the search to find the largest CDs possible, the heuristic function does not account for the 5 alternative restrictions whose size is less than 16 because the size of the 5 alternatives restrictions on the largest CDs on $n \leq 8$ alternatives is larger than or equal to 16 \citep{karpov2024local, leedham2024largest}. We conjecture that this also applied for $n > 8$. Considering that the larger CDs should, in general, contain more number of larger restrictions, we set the weights with increment values $[1, 2, 3, 4, 5]$, i.e $w_{16}=1, w_{17}=2, w_{18}=3, w_{19}=4, w_{20}=5$. 

\begin{figure}[H]
\centering
\begin{subfigure}[b]{0.7\textwidth}
\centering
\includegraphics[width=0.7\textwidth]{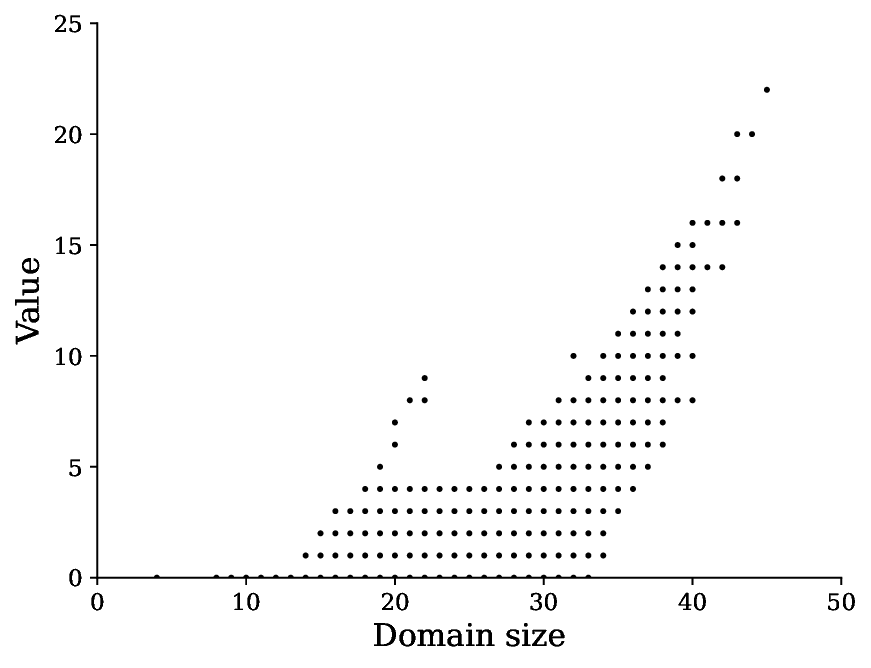}
\caption{The values of all the n=6 non-isomorphic CDs on the heuristic function~(\ref{eq:hf})}
\label{fig:size_value_6}
\end{subfigure}

\begin{subfigure}[b]{0.7\textwidth}
\centering
\includegraphics[width=0.7\textwidth]{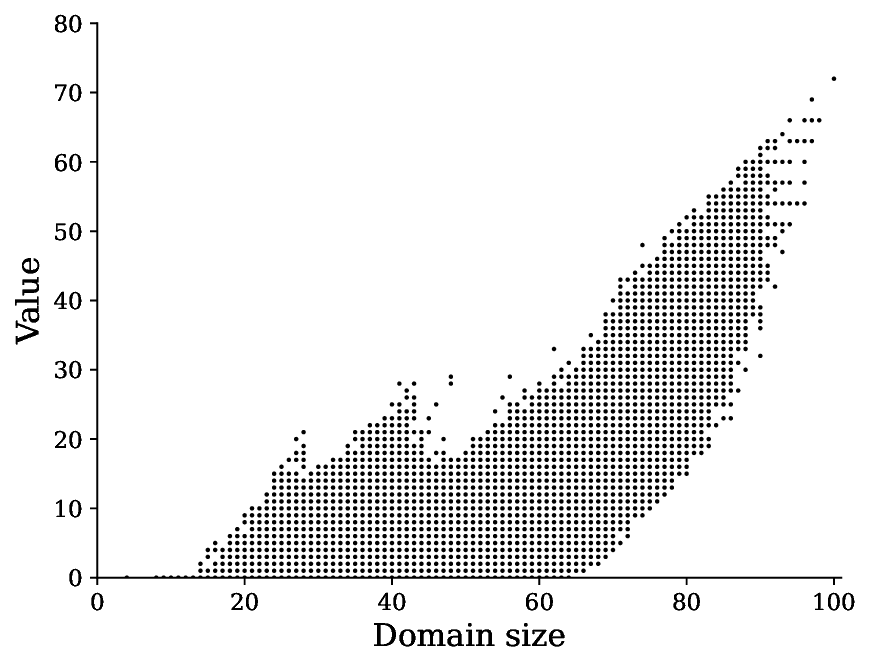}
\caption{The values of some the n=7 non-isomorphic CDs for every size on the heuristic function~(\ref{eq:hf})}
\label{fig:size_value_7}
\end{subfigure}
\captionsetup{width=.9\textwidth}
\caption{The correlation between the sizes of domains and their values evaluated by the value function for six and seven alternative CDs. Plot (a) includes all the six alternative isomorphic CDs; plot (b) uses 100000 randomly sampled $n=7$ CDs for each size if the number of CDs is too many.}
\label{fig:size_value}
\end{figure}

\citet{akello2023condorcet} released two complete sets of unitary non-isomorphic CDs for six and seven alternatives. These CDs enabled us to evaluate the heuristic functions by examining the relationship they give rise to between their sizes and their values. Using these CDs, we show in Fig. \ref{fig:size_value} two scatter plots for six and seven alternatives, demonstrating the relationship between the size of the CDs and their values from our heuristic function. It is conspicuous that for $n=6$ CDs whose size is larger than 28 and for $n=7$ CDs whose size is larger than 44, the heuristic function builds a linear relationship between the size and the values of CDs. The fact that the correlation is not strong on small CDs is not problematic as this issue can be resolved by the prioritisation strategy employed in the search algorithm that could rule them out. Overall, our heuristic function, though not perfect, can serve its purpose well. 

\subsection{The heuristic search algorithm}

In the selection phase of genetic algorithms, solutions in the population are evaluated using a fitness function, which measures how well they solve the problem and the ones that perform have higher fitness are more likely to be selected for the next generation. Owing to its effectiveness, we employ the same strategy in ruling out inferior TRSs that are not likely to end up producing a large CD.


An exhaustive search algorithm that builds an entire search tree begins with a root node corresponding to an empty TRS where the triples are all unassigned. The root node is expanded by adding four child nodes, representing 4 TRSs where the first triple is assigned to one of the four rules, 2N1, 2N3, 1N3, and 3N1. Each child node will then be expanded, and this process continues iteratively until all the triples are assigned. To avoid building an enormous search tree, at each search step our heuristic search algorithm keeps $N$ high-ranking nodes for the next step. The heuristic search algorithm that uses the heuristic function with access to the database is given in Algorithm~\ref{alg:search}.

\begin{algorithm}[H]
\caption{The heuristic search algorithm}\label{alg:search}
\begin{algorithmic}
    \Require{The number of alternatives $n$ where $n \geq 6$ \\
             A set of rules $R$: 2N1, 2N3, 1N3, 3N1 by default \\
             The number of top TRSs kept in the search $N$ where $N \geq 1$\\
             The heuristic function $f$} 
    \Ensure{$\mathbf{TRS}_N$ ($N$ full TRSs with the highest values)} 

    \State Create an empty list $\mathbf{TRS}_N$
    \State Initiate a TRS where all the triples are unassigned and added it to $\mathbf{TRS}_N$
    \State $T \leftarrow$ a list of $m$ triples
    \For {each triple $t$ in $T$}
        \For {each TRS in $\mathbf{TRS}_N$}
            \For {each rule $r$ in $R$}
                \State Assign $r$ to $t$ for TRS
                \State Add TRS to $\mathbf{TRS}_N$
                \State Assign an empty rule to $t$ for TRS
            \EndFor
        \EndFor
       \State Calculate the value for every TRS in $\mathbf{TRS}_N$ using heuristic function $f$
       \State Remove the TRSs from $\mathbf{TRS}_N$  on which the heuristic function returns -1
       \State $\mathbf{TRS}_N$ $\leftarrow$ Top $N$ TRSs with the highest values
    \EndFor
\end{algorithmic}
\end{algorithm}

Choosing a larger value of $N$ may not always lead to better search results. In fact, due to the nonmonotonicity of the heuristic function, increasing $N$ with the same heuristic function can sometimes result in worse search results. This happens because, at the early stage of the search, many TRSs with higher values that lead to smaller CDs outcompete the ones with lower values that could construct larger CDs. Thus, retaining more TRSs could potentially introduce more inferior TRSs.

In such cases, adjusting the parameters in the heuristic function to build a stronger correlation may be necessary. However, our aim is to provide a new paradigm to search for large CDs and we do not presume the heuristic function we developed is optimal. To find better parameters, simply testing out various weight combinations does not help significantly as our experiences indicate. We instead propose to utilise machine learning techniques to find better parameters using the data in \citep{akello2023condorcet}. Another example suggesting the non-monotonic property is that the search algorithm using a heuristic function calculating the value from restrictions on 6 alternatives did not find any larger CDs than the Fishburn domains. Here, as it is prohibitively impractical to build the dataset for 6 alternatives, we computed the largest possible restriction sizes using the process described in Algorithm~\ref{alg:database}.

Searching for large CDs entails a significant amount of computation, especially on CDs with many alternatives. One of the advantages of our algorithms is that the computation needed to evaluate each partial CD by the value function is much less than calculating its actual size, and the overhead of working with a larger number of alternatives only increases slightly. While our search algorithms have discovered new unknown large CDs for 10 and 11 alternatives, we conjecture there are larger CDs for $n=9$ that our search algorithms have not found, indicating that there is still some room for further improvement. The heuristic function we developed, based on the 5 alternative restriction sizes, is practically accurate, but it is by no means perfect. Its success shows a strong correlation between the actual CD size and its corresponding subset sizes, and exploiting this feature in more innovative ways opens a door for future research. At least, given the current construction, many other weight combinations can also be tested out. 

\section{Experiments}
\label{sec:experiments}

All our experiments were built on top of the CDL library \citep{zhou2023cdl} which provides a wide range of functionalities for CD-related calculations. This library was initially developed to support this work. It has underpinned some research studies \citep{karpov2023set, akello2023condorcet, karpov2024local, markstrom2024arrow} since its initial release due to its flexibility and fast execution speed. The fast computation enabled by this library contributes significantly to the success of our algorithm. Especially, the \texttt{subset\_states} function it provided, which converts a TRS to its subset states for $n\geq6$ alternatives, facilitated the implementation of our search algorithms. 

Our search algorithm found the largest CD for $n=6$ within a second, as shown in Table~\ref{tab:size}. As the number of alternatives increased, the time it took to find a large CD skyrocketed not only because the search space was growing exponentially but because more high-value nodes were needed to participate in the search to get good results. It is worth noting that this heuristic search algorithm has been integrated into the CDL library. The codes are available at the CDL GitHub repository\footnote{GitHub. \url{https://github.com/sagebei/cdl/tree/main/python}}.

\begin{table}[H]
\captionsetup{width=.55\textwidth}
\caption{The time consumption of our method on rediscovering the known largest CD for $n\leq9$ on a Macbook Air with 1.1 GHz Dual-Core Intel Core i3 and 8 Gigabyte memory.}
\centering
\small
\begin{tabular}{ccc}
\toprule
\textbf{n} & \textbf{Size of largest CDs found} & \textbf{Time Consumption}\\
\midrule

6 & 45 & 0.605 second\\
7 & 100 & 1.31 second\\
8 & 222 & 93 seconds\\
9 & 488 & 31 minutes\\

\bottomrule
\end{tabular}
\label{tab:size}
\end{table}

On more than nine alternatives, the order of the triples by which they are assigned with a rule is crucial and has a non-negligible impact on the CDs found by our search algorithm. A well-chosen ordering helps cut off unfruitful branches at the early search stage. To verify this, we developed an approach that selects the next triple to be assigned a rule dynamically, aiming to find the triple among all the unassigned ones that reduce the size of the CD the most once assigned. This approach records the size of a partial CD for each unassigned triple when assigned with one of the 4 rules that maximize the size of this partial CD and choose the one with the lowest value. Dynamically selecting the next triple gives good results, but it requires a large amount of calculation to access all the unassigned triples, and it is prohibitively expensive to employ this strategy when working on $n\ge8$ alternatives.

We experimented with different static triple orders where the triples were listed randomly or sorted, and the order was fixed throughout the search once determined. One of our special static orders based on various mathematical heuristics and proprieties stood out and led to the discovery of record-breaking CDs. The ordering of the triples we used is a type of static ordering and is defined such that triple $(x_1,y_1,z_1)$ is before $(x_2,y_2,z_2)$ if $x_1<x_2$ or $(x_1=x_2$ and $z_1<z_2)$ or $(x_1=x_2$ and $z_1=z_2$ and $y_1<y_2)$. This choice of triple order is a contributing factor that enables our method to find the existing large CDs constructed by alternating schemes quickly. We call this ordering RZ-ordering, where RZ is the authors' initials. 

One major benefit brought by RZ-order is that the order of the list of triples in a TRS on $n\geq6$ alternatives when restricted to five alternatives remains unchanged, which is to say that when the triples in the five alternatives database are in RZ-order, for $n\geq6$ alternatives the restricted triples to five alternatives are also in RZ-order, meaning that they can be looked up in that database without reordering. However, for other orders of triples that do not hold this property, like the lexicographic order, many of the orders of triples restricted to five alternatives are not in lexicographic order. They have to be reordered before being able to be looked up in the dataset where triples are also in lexicographic order, incurring extra cost of computation. 

In the experiments where the new CDs are found, we kept the top $N=200000$  and  $N=100000$ high-value nodes for 10 and 11 alternatives, respectively. Table~\ref{tab:results} displays the size and their counts of the CDs that are larger than the Fishburn domains which are the largest known CDs on 10 and 11 alternatives. The largest CDs we found on 10 alternatives have size 1082, while the size of other smaller CDs ranges from 1070 to 1079. On 11 alternatives, there are CDs of 5 different sizes larger than the Fishburn domain discovered by the search algorithm, with the largest one being 2349.  

It is sensible to claim that the larger the size of CDs, the harder they are to be discovered. However, our method found more (isomorphic) CDs of size 1072 than those of size 1071 for 10 alternatives and more CDs of size 2337 than those of size 2334 for 11 alternatives. This reflects one of the challenges mentioned above in the search where designing a successful heuristic function that prioritizes the large CD is challenging. It also insinuates that the heuristic function we crafted, although with significant efforts, is imperfect. We see it as an open research problem to design an efficient and effective heuristic function that distinguishes the CDs of different sizes with a large margin such that the unpromising partial CDs can be ruled out with high confidence. 

\begin{table}[H]
\captionsetup{width=.7\textwidth}
\caption{The sizes and counts of new large CDs on 10 and 11 alternatives, discovered by the heuristic search algorithm.}
    \centering
    \small
    \begin{tabular}{ccccccccc}
    \toprule
    \textbf{n} & \multicolumn{8}{c}{\textbf{New large CD sizes and their count}}\\

    \midrule
    \multirow{2}{*}{10} & size & 1070 & 1071 & 1072 & 1074 & 1078 & 1079 & 1082 \\
    & count & 80 & 29 & 70 & 37 & 13 & 8 & 5\\ 

    \midrule
    \multirow{2}{*}{11} & size & 2328 & 2329 & 2334 & 2337 & 2349\\
    & count & 330 & 16 & 8 & 67 & 122\\ 
    
    \bottomrule
\end{tabular}
    
    \label{tab:results}
\end{table}

This search algorithm, though relying on access to a database, can be scaled to run on thousands of CPU cores on parallel, each of which independently starts a search on a partial CD. We experimented with this technique by starting 1000 searches on 1000 CPU cores on QMUL HPC machine \citep{king2021apocrita}, and found CDs whose size is 224 for eight alternatives, the largest CD for $n=8$ \citep{leedham2024largest}. 

The whole search was split into two stages. In the first stage, the search began with a TRS with no rule assigned and stopped after a given triple (the 15th triple in our experiment) in TRSs was assigned, resulting in a large collection of TRSs where the top triples above that triple (the first 15 triples in our case) were all assigned. The first stage ended with evenly splitting the list of TRSs into 1000 chunks. They were shuffled before splitting to ensure that the TRSs in one chunk were not similar. In the second stage, we resumed the search on 1000 CPU cores with the $N$, the number of high-value CDs kept being 10000, each of which searched for a chunk independently. When they were finished, the size of all the CDs found by the search was calculated, and we found 17 CDs of size 224. 

Paralleling the search algorithm in this way undoubtedly accelerates the computation by engaging in many CPU cores, but it has one drawback. As the list of TRSs is segregated into isolated blocks, and the search is conducted within each block, each block is only aware of the values of the TRSs in itself. 

\section{Analysis on discovered new large CDs}
\label{sec:analysis}

As the restrictions of a CD reveal its inner local structure and their application plays a central role in the heuristic function employed in the search algorithm, we use the size of restrictions and their counts as a measure of the CD structure. On 11 alternatives, the search algorithms found five CDs of different sizes larger than the Fishburn domain. Table~\ref{tab:11_cd_subsets} shows that each of these large CDs when restricted to 10 alternative subsets, produces at least one restriction larger than the Fishburn domain. Still, they were all found by our method, indicating our method is consistent across different numbers of alternatives. It is also worth pointing out that none of the restrictions is identical to the Fishburn domain, showing that the inner structure of these new CDs differs fundamentally from that of the Fishburn domain. 

\begin{table}[H]
\captionsetup{width=.95\textwidth}
\caption{The 10 alternative restriction sizes of these $n=11$ CDs larger than the Fishburn domain. Each contains at least one n=10 CD larger than the Fishburn domain when restricted to some 10 alternatives. These CD sizes are highlighted in bold.}
\centering
\small
\begin{tabular}{ccccccccccc}
\toprule
\textbf{CD size for n=11}  & \multicolumn{10}{c}{\textbf{Restriction sizes on 10 alternatives}} \\
\midrule
2328  & 1003 & 1005 & 1022 & 1043 & 1053 & \textbf{1072} & \textbf{1074} & \textbf{1079}\\
2329 & 1013 & 1028 & 1039 & 1040 & 1041 & 1044 & 1057  & 1058 & 1063 & \textbf{1079}\\
2334 & 1021 & 1030 & 1035 & 1039 & 1043 & 1057 & 1063 & 1064 & \textbf{1078}\\
2337 & 1017 & 1022 & 1031 & 1045 & 1053 & 1062 & \textbf{1072} & \textbf{1074}\\
2349  & 1021 & 1026 & 1035 & 1045 & 1053 & 1068 & \textbf{1074} & \textbf{1078}\\
 \bottomrule
\end{tabular}
\label{tab:11_cd_subsets}
\end{table}

On 10 and 11 alternatives, the Fishburn domains had the largest size before our results. Here we compare the largest CDs found by the search algorithms with them.
We calculated the restriction sizes for the largest CD we found on $n=10$ and $11$ alternatives when restricted to a range of 4 to $n-1$ alternatives. Their sizes and counts are listed in Table~\ref{tab:subset_sizes_10} and Table~\ref{tab:subset_sizes_11} for 10 and 11 alternatives respectively. At least one of the restrictions on the 10 alternative Fishburn domain is also a Fishburn domain. However, for the largest CD on 10 alternatives our algorithm found, the size of the largest restriction on nine alternatives is 485, which is smaller than the Fishburn domain. We thus speculate that the chance that it is not the largest is high. It provides an improved low bound, but there is no evidence showing that it is the high bound on the domain size. 

Furthermore, the restrictions from Fishburn domains generally have a larger number of large CDs than the ones from the CD of size 1082. For example, when restricted to eight alternatives, only one restriction is of size 222, in contrast to the ones from the Fishburn domain, which contain 15 restrictions whose size is 222. This shows that larger CDs do not always produce larger restrictions or a higher number of larger restrictions compared to those from smaller CDs. This aligns with the conclusion in \citet{karpov2024local} that suggests CDs that have the largest restriction might not be the largest. 

\begin{table}[ht]
    \centering
    \small
    \captionsetup{width=0.9\linewidth}
    \caption{The restriction sizes and their count for the 10 alternative CD of size 1082 when restricted to 4 to 9 alternative subsets. }
    
    \begin{tabular}{clccccccccccc}
    \toprule
    \textbf{Subset n} & \multicolumn{12}{c}{\textbf{Restriction sizes and their count}}\\

    \midrule
    \multirow{2}{*}{4} & size & \textbf{8} & 9 \\
    & count & 90 & 120\\ 

    \midrule
    \multirow{2}{*}{5} & size & \textbf{16} & 17 & 18 & 19 & 20\\
    & count & 6 & 30 & 6 & 162 & 48\\ 
    
    \midrule
    \multirow{2}{*}{6} & size & 36 & 39 & 40 & 41 & 42 & 43 & 44 & 45 \\
    & count & 6 & 6 & 2 & 44 & 86 & 40 & 6 & 20\\ 

 \midrule
    \multirow{2}{*}{7} & size & 87 & 89 & 91 & 92 & 93 & 94 & 95 & 96 & 97 & 98 & 100 \\
& count & 6 & 2 & 4 & 16 & 22 & 18 & 8 & 16 & 6 & 16 & 6 \\

     \midrule
    \multirow{2}{*}{8} & size & 200 & 204 & 205 & 209 & 211 & 212 & 214 & 216 & 218 & 219 & 222 \\
&count& 2 & 4 & 4 & 4 & 3 & 2 & 7 & 10 & 4 & 4 & 1 \\

    \midrule
    \multirow{2}{*}{9} & size & 473 & 481 & 485 \\
    & count & 4 & 2 & 4 \\
    
    \bottomrule
\end{tabular}
    
\label{tab:subset_sizes_10}
\end{table}

We observed a similar pattern with the restrictions on the 11 alternative CD of size 2349. As shown in Table \ref{tab:subset_sizes_11}, some of these restrictions on the 5, 6, and 7 alternatives, have the same size as the the Fishburn domains. But it also has relatively small restrictions, especially on seven alternatives where the size of the smallest restriction is 74. All the restrictions on 9 alternatives are smaller than Fishburn domains.  

\begin{table}[ht]
    \centering
    \small
    \captionsetup{width=.88\textwidth}
    \caption{The restriction sizes and their count for the 11 alternative CD of size 2349 when restricted to 4 to 10 alternative subsets.}
    \label{tab:subset_sizes_11}
    
    \setlength{\tabcolsep}{4pt}
    \begin{tabular}{clcccccccccccc}
    \toprule
    \textbf{Subset n} & \multicolumn{13}{c}{\textbf{Restriction sizes and their count}}\\

    \midrule
    \multirow{2}{*}{4} & size & \textbf{8} & 9 \\
    & count & 143 & 187\\ 

    \midrule
    \multirow{2}{*}{5} & size & \textbf{16} & 17 & 18 & 19 & 20\\
    & count & 28 & 46 & 5 & 279 & 104\\ 

     \midrule
    \multirow{2}{*}{6} & size & \textbf{32} & 35 & 36 & 39 & 40 & 41 & 42 & 43 & 44 & 45\\
    & count & 3 & 10 & 9 & 35 & 2 & 73 & 164 & 79 & 35 & 52\\
    
     \midrule
    \multirow{4}{*}{7} & size & 74 & 79 & 85 & 86 & 87 & 88 & 89 & 91 & 92 & 93 & 94 & 95  \\
    & count & 3 & 3 & 12 & 3 & 9 & 3 & 15 & 19 & 44 & 39 & 12 & 16   \\
    & size & 96& 97 & 98 & 100 \\
    & count& 39& 52 & 42 & 19 \\
    
    \midrule
    \multirow{4}{*}{8} & size & 179 & 184 & 187 & 192 & 194 & 196 & 200 & 201 & 202 & 204 & 205 & 207  \\
    & count &3 & 1 & 1 & 1 & 3 & 8 & 3 & 2 & 8 & 6 & 2 & 7 \\
    & size & 209 & 210 & 211 & 212& 213 & 214 & 215 & 216 & 217 & 218 & 219 & 222\\
    & count& 10 & 3 & 12 & 17 & 11 & 2 & 8 & 13 & 2 & 30 & 8 & 4\\
    
    \midrule
    \multirow{4}{*}{9} & size & 415 & 426 & 431 & 448 & 451 & 452 & 457 & 460 & 466 & 468 & 470 & 473 \\
    & count & 1 & 1 & 1 & 2 & 2 & 1 & 1 & 1 & 4 & 2 & 2 & 4 \\
    & size& 475 & 478 & 479 & 480 & 481 & 484 & 485\\
    & count & 3 & 4 & 3 & 4  & 11 & 4 & 4\\
    \midrule
    \multirow{2}{*}{10} & size & 1021 & 1026 & 1035 & 1045 & 1053 & 1068 & \textbf{1074} & \textbf{1078}\\
    & count & 1 & 1 & 1 & 2 & 1 & 1 & 2 & 2 \\

    \bottomrule
\end{tabular}
\end{table}

We also noticed that these new large CDs have some relatively small restrictions of sizes 8 and 16 when restricted to four and five alternatives, respectively. This was not a surprise to us as it generally follows Ramsey's Theorem \citep{ramsey}, which in the context of graph colouring states that given a positive integer $k$ there exists a complete graph with $n (n > k)$ coloured edges where a subset of $k$ connected edges have the same colour. We will show that Ramsey's Theorem also applies to analysing the pattern of the rules assigned to the triples for the restrictions of a CD. We state that:

\begin{Cor}
For any positive integer $k$, there exists a positive integer $n (n>k)$, such that any CD on $n$ alternatives has a restriction to $k$ alternatives where the rules on them are the same.
\end{Cor}

\begin{proofoutline} 
The proof comes with a direct application of Ramsey's Theorem if we consider the triples as edges of a graph and the rule assigned to them as their colour. 
\end{proofoutline}

\begin{Cor}
For any positive integer $k$, there exists a positive integer $n (n>k)$, such that any CD on $n$ alternatives has a restriction to $k$ alternatives of size $2^{k-1}$. 
\end{Cor}
\begin{proofoutline} 
A CD on $n$ alternatives created by a list of triples where all the rules assigned to them are the same, i.e. all 1N3, 3N1, 2N3, 2N1, 1N2 or 3N2 has size $2^{n-1}$ \citep{raynaud1981paradoxical}. 
\end{proofoutline}

Our finding here shows that on this $n=10$ CD there exist restrictions on 4 and 5 alternatives with size 8 and 16 respectively, meaning that for $k$ values of 4 and 5, $n$ value of 10 already satisfies the condition. On the 11 alternatives CD of size 2349, the condition for $k=6$ is satisfied on the largest CD we discovered on $n=11$ alternatives where some of the restrictions on 6 alternatives have size 32.

\section{Conclusion}
\label{sec:conclusion}

In this paper, we present a heuristic search algorithm capable of generating large CDs on any number of alternatives, which also has the potential to discover unknown larger CDs. This is evidenced by the fact that the large CDs it found on the 10 and 11 alternatives improved the existing lower bounds. The success of our algorithm lies chiefly in its utilisation of human domain knowledge on this subject which existing learning algorithms for general combinatorial optimisation problems have no apparent way to assimilate. Our results were obtained without applying any statistic planning approaches, like Monte Carlo Tree Search, but by traversing the tree using a purely manually designed heuristic function to guide the search. The way we traversed the search tree only used a pure heuristic function but did not explicitly use any reinforcement learning techniques like Q-learning \citep{watkins1992q}. Despite having access to limited computational resources, our algorithm found many new large CDs. 

Although the application of the database does cut away a large part of the search tree at a cheap cost, it has certain limitations. With a reasonable amount of computations, the algorithm could not find a new lower bound for CDs on 12 alternatives, suggesting that the CDs with a larger number of alternatives have more non-local implications, and the upper bound given by the database will more often differ from the actual size of the local restriction. 

Finding the largest CDs is a challenging yet unsolved problem. Many existing algorithms can be applied to finding large CDs. We tested various standard approaches, including reinforcement learning algorithms, evolutionary algorithms and local search algorithms. In our preliminary experiments, they only managed to find local maxima solutions and had great difficulties in finding the largest known domains for low values of n, such as $n=7$ and $n=8$. The new CDs we report suggest that there may be more complex mathematical patterns underlying the CDs than previously thought. Our results indicate that the maximal size of CDs on $n\geq8$ alternatives can serve as a benchmark for heuristic search algorithms. The problems are challenging because the search space is enormous on large alternatives, and there are no obvious, simple criteria for searching. Still, at the same time, there are a lot of structures in the CD as the restrictions to each subset of alternatives also form a CD. An efficient search algorithm should be able to exploit this rich underlying structure to guide the search.

\section*{Declarations}
\textbf{Conflict of interest} The authors have no relevant financial or non-financial interests that could potentially compromise the integrity, impartiality, or credibility of the research presented in this paper.

\section*{Acknowledgements}

Bei Zhou was funded by the China Scholarship Council (CSC). This research utilised Queen Mary's Apocrita HPC facility \citep{king2021apocrita}, supported by QMUL Research-IT.

\bibliographystyle{plainnat}
\bibliography{references}

\section*{Appendix}

\subsection*{The triples and their rule for the n=10 CD of size 1082}

\begin{table}[H]
\centering
\small
\setlength\extrarowheight{-3pt}
\captionsetup{width=1.0\linewidth}
\caption{The TRS that led to the CD of size 1082 on 10 alternatives. The ordering of the triples displayed is another static ordering that differs from the ordering used in the search algorithm and is defined such that $(x_1,y_1,z_1)$ is before $(x_2,y_2,z_2)$ if $x_1 \leq x_2$ and $y_1 \leq y_2$ and $z_1 \leq z_2$ out of consideration for the ease of looking up the table manually. There are 1082 isomorphic versions of this CD. The 10 alternative CD built by these rules is presented in Sect.~\ref{sec:domain}.}
\begin{tabular}{lc}
\toprule
\textbf{Triples} & \textbf{Rules}\\
\midrule

(1, 2, 3) & 2N3 \\
(1, 2, 4) & 2N3 \\
(1, 2, 5) & 1N3 \\
(1, 2, 6) & 1N3 \\
(1, 2, 7) & 1N3 \\
(1, 2, 8) & 3N1 \\
(1, 2, 9) & 2N3 \\
(1, 2, 10) & 2N3 \\
(1, 3, 4) & 2N3 \\
(1, 3, 5) & 1N3 \\
(1, 3, 6) & 1N3 \\
(1, 3, 7) & 1N3 \\
(1, 3, 8) & 3N1 \\
(1, 3, 9) & 3N1 \\
(1, 3, 10) & 3N1 \\
(1, 4, 5) & 1N3 \\
(1, 4, 6) & 1N3 \\
(1, 4, 7) & 1N3 \\
(1, 4, 8) & 3N1 \\
(1, 4, 9) & 3N1 \\
(1, 4, 10) & 3N1 \\
(1, 5, 6) & 3N1 \\
(1, 5, 7) & 3N1 \\
(1, 5, 8) & 1N3 \\
(1, 5, 9) & 1N3 \\
(1, 5, 10) & 1N3 \\
(1, 6, 7) & 3N1 \\
(1, 6, 8) & 1N3 \\
(1, 6, 9) & 1N3 \\
(1, 6, 10) & 1N3 \\
\bottomrule
\end{tabular}
\hspace{0em}
\begin{tabular}{lc}
\toprule
\textbf{Triples} & \textbf{Rules}\\
\midrule
(1, 7, 8) & 1N3 \\
(1, 7, 9) & 1N3 \\
(1, 7, 10) & 1N3 \\
(1, 8, 9) & 2N1 \\
(1, 8, 10) & 2N1 \\
(1, 9, 10) & 3N1 \\
(2, 3, 4) & 2N3 \\
(2, 3, 5) & 1N3 \\
(2, 3, 6) & 1N3 \\
(2, 3, 7) & 1N3 \\
(2, 3, 8) & 1N3 \\
(2, 3, 9) & 3N1 \\
(2, 3, 10) & 3N1 \\
(2, 4, 5) & 1N3 \\
(2, 4, 6) & 1N3 \\
(2, 4, 7) & 1N3 \\
(2, 4, 8) & 1N3 \\
(2, 4, 9) & 3N1 \\
(2, 4, 10) & 3N1 \\
(2, 5, 6) & 3N1 \\
(2, 5, 7) & 3N1 \\
(2, 5, 8) & 3N1 \\
(2, 5, 9) & 1N3 \\
(2, 5, 10) & 1N3 \\
(2, 6, 7) & 3N1 \\
(2, 6, 8) & 3N1 \\
(2, 6, 9) & 1N3 \\
(2, 6, 10) & 1N3 \\
(2, 7, 8) & 3N1 \\
(2, 7, 9) & 1N3 \\
\bottomrule
\end{tabular}
\hspace{0em}
\begin{tabular}{lc}
\toprule
\textbf{Triples} & \textbf{Rules}\\
\midrule
(2, 7, 10) & 1N3 \\
(2, 8, 9) & 1N3 \\
(2, 8, 10) & 1N3 \\
(2, 9, 10) & 3N1 \\
(3, 4, 5) & 1N3 \\
(3, 4, 6) & 1N3 \\
(3, 4, 7) & 1N3 \\
(3, 4, 8) & 1N3 \\
(3, 4, 9) & 1N3 \\
(3, 4, 10) & 1N3 \\
(3, 5, 6) & 3N1 \\
(3, 5, 7) & 3N1 \\
(3, 5, 8) & 3N1 \\
(3, 5, 9) & 1N3 \\
(3, 5, 10) & 3N1 \\
(3, 6, 7) & 3N1 \\
(3, 6, 8) & 3N1 \\
(3, 6, 9) & 1N3 \\
(3, 6, 10) & 3N1 \\
(3, 7, 8) & 3N1 \\
(3, 7, 9) & 1N3 \\
(3, 7, 10) & 3N1 \\
(3, 8, 9) & 1N3 \\
(3, 8, 10) & 3N1 \\
(3, 9, 10) & 1N3 \\
(4, 5, 6) & 3N1 \\
(4, 5, 7) & 3N1 \\
(4, 5, 8) & 3N1 \\
(4, 5, 9) & 1N3 \\
(4, 5, 10) & 3N1 \\
\bottomrule
\end{tabular}
\hspace{0em}
\begin{tabular}{lc}
\toprule
\textbf{Triples} & \textbf{Rules}\\
\midrule
(4, 6, 7) & 3N1 \\
(4, 6, 8) & 3N1 \\
(4, 6, 9) & 1N3 \\
(4, 6, 10) & 3N1 \\
(4, 7, 8) & 3N1 \\
(4, 7, 9) & 1N3 \\
(4, 7, 10) & 3N1 \\
(4, 8, 9) & 1N3 \\
(4, 8, 10) & 3N1 \\
(4, 9, 10) & 1N3 \\
(5, 6, 7) & 3N1 \\
(5, 6, 8) & 1N3 \\
(5, 6, 9) & 2N1 \\
(5, 6, 10) & 1N3 \\
(5, 7, 8) & 1N3 \\
(5, 7, 9) & 2N1 \\
(5, 7, 10) & 1N3 \\
(5, 8, 9) & 2N1 \\
(5, 8, 10) & 2N1 \\
(5, 9, 10) & 3N1 \\
(6, 7, 8) & 2N1 \\
(6, 7, 9) & 2N1 \\
(6, 7, 10) & 2N1 \\
(6, 8, 9) & 2N1 \\
(6, 8, 10) & 2N1 \\
(6, 9, 10) & 3N1 \\
(7, 8, 9) & 2N1 \\
(7, 8, 10) & 2N1 \\
(7, 9, 10) & 3N1 \\
(8, 9, 10) & 3N1 \\

\bottomrule
\end{tabular}
\label{table:triples_rules_10}
\end{table}

\subsection*{The triples and their rule for the n=11 CD of size 2349}
\label{sec:11}

\begin{table}[H]
\centering
\small
\setlength\extrarowheight{-3pt}
\captionsetup{width=.88\textwidth}
\caption{The TRS that constructed CD of size 2349 on 11 alternatives. The 11 alternative CD built by these rules is presented in Sect.~\ref{sec:domain}.}
\begin{tabular}{lc}
\toprule
\textbf{Triples} & \textbf{Rules}\\
\midrule

(1, 2, 3) & 1N3 \\
(1, 2, 4) & 2N3 \\
(1, 2, 5) & 2N3 \\
(1, 2, 6) & 2N3 \\
(1, 2, 7) & 3N1 \\
(1, 2, 8) & 1N3 \\
(1, 2, 9) & 1N3 \\
(1, 2, 10) & 1N3 \\
(1, 2, 11) & 1N3 \\
(1, 3, 4) & 1N3 \\
(1, 3, 5) & 1N3 \\
(1, 3, 6) & 1N3 \\
(1, 3, 7) & 1N3 \\
(1, 3, 8) & 3N1 \\
(1, 3, 9) & 2N3 \\
(1, 3, 10) & 3N1 \\
(1, 3, 11) & 2N3 \\
(1, 4, 5) & 3N1 \\
(1, 4, 6) & 3N1 \\
(1, 4, 7) & 3N1 \\
(1, 4, 8) & 1N3 \\
(1, 4, 9) & 1N3 \\
(1, 4, 10) & 1N3 \\
(1, 4, 11) & 1N3 \\
(1, 5, 6) & 3N1 \\
(1, 5, 7) & 3N1 \\
(1, 5, 8) & 1N3 \\
(1, 5, 9) & 1N3 \\
(1, 5, 10) & 1N3 \\
(1, 5, 11) & 1N3 \\
(1, 6, 7) & 3N1 \\
(1, 6, 8) & 1N3 \\
(1, 6, 9) & 1N3 \\
(1, 6, 10) & 1N3 \\
(1, 6, 11) & 1N3 \\
(1, 7, 8) & 1N3 \\
(1, 7, 9) & 1N3 \\
(1, 7, 10) & 1N3 \\
(1, 7, 11) & 1N3 \\
(1, 8, 9) & 2N1 \\
(1, 8, 10) & 2N1 \\
(1, 8, 11) & 2N1 \\
\bottomrule
\end{tabular}
\hspace{0.1em}
\begin{tabular}{lc}
\toprule
\textbf{Triples} & \textbf{Rules}\\
\midrule
(1, 9, 10) & 3N1 \\
(1, 9, 11) & 3N1 \\
(1, 10, 11) & 2N1 \\
(2, 3, 4) & 1N3 \\
(2, 3, 5) & 1N3 \\
(2, 3, 6) & 1N3 \\
(2, 3, 7) & 3N1 \\
(2, 3, 8) & 3N1 \\
(2, 3, 9) & 2N3 \\
(2, 3, 10) & 3N1 \\
(2, 3, 11) & 2N3 \\
(2, 4, 5) & 3N1 \\
(2, 4, 6) & 3N1 \\
(2, 4, 7) & 1N3 \\
(2, 4, 8) & 1N3 \\
(2, 4, 9) & 1N3 \\
(2, 4, 10) & 1N3 \\
(2, 4, 11) & 1N3 \\
(2, 5, 6) & 3N1 \\
(2, 5, 7) & 1N3 \\
(2, 5, 8) & 1N3 \\
(2, 5, 9) & 1N3 \\
(2, 5, 10) & 1N3 \\
(2, 5, 11) & 1N3 \\
(2, 6, 7) & 1N3 \\
(2, 6, 8) & 1N3 \\
(2, 6, 9) & 1N3 \\
(2, 6, 10) & 1N3 \\
(2, 6, 11) & 1N3 \\
(2, 7, 8) & 2N1 \\
(2, 7, 9) & 2N1 \\
(2, 7, 10) & 2N1 \\
(2, 7, 11) & 2N1 \\
(2, 8, 9) & 2N1 \\
(2, 8, 10) & 2N1 \\
(2, 8, 11) & 2N1 \\
(2, 9, 10) & 3N1 \\
(2, 9, 11) & 3N1 \\
(2, 10, 11) & 2N1 \\
(3, 4, 5) & 3N1 \\
(3, 4, 6) & 3N1 \\
(3, 4, 7) & 3N1 \\
\bottomrule
\end{tabular}
\hspace{0.1em}
\begin{tabular}{lc}
\toprule
\textbf{Triples} & \textbf{Rules}\\
\midrule
(3, 4, 8) & 3N1 \\
(3, 4, 9) & 1N3 \\
(3, 4, 10) & 3N1 \\
(3, 4, 11) & 1N3 \\
(3, 5, 6) & 3N1 \\
(3, 5, 7) & 3N1 \\
(3, 5, 8) & 3N1 \\
(3, 5, 9) & 1N3 \\
(3, 5, 10) & 3N1 \\
(3, 5, 11) & 1N3 \\
(3, 6, 7) & 3N1 \\
(3, 6, 8) & 3N1 \\
(3, 6, 9) & 1N3 \\
(3, 6, 10) & 3N1 \\
(3, 6, 11) & 1N3 \\
(3, 7, 8) & 3N1 \\
(3, 7, 9) & 1N3 \\
(3, 7, 10) & 3N1 \\
(3, 7, 11) & 1N3 \\
(3, 8, 9) & 1N3 \\
(3, 8, 10) & 3N1 \\
(3, 8, 11) & 1N3 \\
(3, 9, 10) & 1N3 \\
(3, 9, 11) & 3N1 \\
(3, 10, 11) & 1N3 \\
(4, 5, 6) & 3N1 \\
(4, 5, 7) & 1N3 \\
(4, 5, 8) & 1N3 \\
(4, 5, 9) & 2N1 \\
(4, 5, 10) & 1N3 \\
(4, 5, 11) & 2N1 \\
(4, 6, 7) & 1N3 \\
(4, 6, 8) & 1N3 \\
(4, 6, 9) & 2N1 \\
(4, 6, 10) & 1N3 \\
(4, 6, 11) & 2N1 \\
(4, 7, 8) & 2N1 \\
(4, 7, 9) & 2N1 \\
(4, 7, 10) & 2N1 \\
(4, 7, 11) & 2N1 \\
(4, 8, 9) & 2N1 \\
(4, 8, 10) & 2N1 \\
\bottomrule
\end{tabular}
\hspace{0.1em}
\begin{tabular}{lc}
\toprule
\textbf{Triples} & \textbf{Rules}\\
\midrule
(4, 8, 11) & 2N1 \\
(4, 9, 10) & 3N1 \\
(4, 9, 11) & 2N3 \\
(4, 10, 11) & 2N1 \\
(5, 6, 7) & 2N1 \\
(5, 6, 8) & 2N1 \\
(5, 6, 9) & 2N1 \\
(5, 6, 10) & 2N1 \\
(5, 6, 11) & 2N1 \\
(5, 7, 8) & 2N1 \\
(5, 7, 9) & 2N1 \\
(5, 7, 10) & 2N1 \\
(5, 7, 11) & 2N1 \\
(5, 8, 9) & 2N1 \\
(5, 8, 10) & 2N1 \\
(5, 8, 11) & 2N1 \\
(5, 9, 10) & 3N1 \\
(5, 9, 11) & 2N3 \\
(5, 10, 11) & 2N1 \\
(6, 7, 8) & 2N1 \\
(6, 7, 9) & 2N1 \\
(6, 7, 10) & 2N1 \\
(6, 7, 11) & 2N1 \\
(6, 8, 9) & 2N1 \\
(6, 8, 10) & 2N1 \\
(6, 8, 11) & 2N1 \\
(6, 9, 10) & 3N1 \\
(6, 9, 11) & 2N3 \\
(6, 10, 11) & 2N1 \\
(7, 8, 9) & 2N1 \\
(7, 8, 10) & 2N1 \\
(7, 8, 11) & 2N1 \\
(7, 9, 10) & 3N1 \\
(7, 9, 11) & 2N3 \\
(7, 10, 11) & 2N1 \\
(8, 9, 10) & 3N1 \\
(8, 9, 11) & 2N3 \\
(8, 10, 11) & 2N1 \\
(9, 10, 11) & 1N3 \\

\bottomrule
\vspace{2.5em}
\end{tabular}

\label{table:triples_rules_11}
\end{table}

\subsection*{The Condorcet domains}
\label{sec:domain}

The large Condorcet domains we discovered for 10 and 11 alternatives are demonstrated here. We use hexadecimal numbers A and B to represent the domain's decimal numbers 10 and 11. 

\subsubsection*{The Condorcet domain of size 1082 on 10 alternatives}
\label{sec:cd_10}

\begin{spacing}{0.8}
\footnotesize
\noindent 
123456789A
12345678A9
123456798A
12345679A8
123456978A
12345697A8
1234569A78
1234569A87
123459678A
12345967A8
1234596A78
1234596A87
123459A678
123459A687
123459A867
123459A876
123495678A
12349567A8
1234956A78
1234956A87
123495A678
123495A687
123495A867
123495A876
123496578A
12349657A8
1234965A78
1234965A87
123496758A
12349675A8
12349A5678
12349A5687
12349A5867
12349A5876
12349A8567
12349A8576
123546789A
12354678A9
123546798A
12354679A8
123546978A
12354697A8
1235469A78
1235469A87
123549678A
12354967A8
1235496A78
1235496A87
123549A678
123549A687
123549A867
123549A876
123564789A
12356478A9
123564798A
12356479A8
123564978A
12356497A8
1235649A78
1235649A87
123567489A
12356748A9
123567498A
12356749A8
123567849A
12356784A9
1235678A49
123945678A
12394567A8
1239456A78
1239456A87
123945A678
123945A687
123945A867
123945A876
123946578A
12394657A8
1239465A78
1239465A87
123946758A
12394675A8
12394A5678
12394A5687
12394A5867
12394A5876
12394A8567
12394A8576
125346789A
12534678A9
125346798A
12534679A8
125346978A
12534697A8
1253469A78
1253469A87
125349678A
12534967A8
1253496A78
1253496A87
125349A678
125349A687
125349A867
125349A876
125364789A
12536478A9
125364798A
12536479A8
125364978A
12536497A8
1253649A78
1253649A87
125367489A
12536748A9
125367498A
12536749A8
125367849A
12536784A9
1253678A49
125634789A
12563478A9
125634798A
12563479A8
125634978A
12563497A8
1256349A78
1256349A87
125637489A
12563748A9
125637498A
12563749A8
125637849A
12563784A9
1256378A49
125673489A
12567348A9
125673498A
12567349A8
125673849A
12567384A9
1256738A49
125678349A
12567834A9
1256783A49
125678A349
129345678A
12934567A8
1293456A78
1293456A87
129345A678
129345A687
129345A867
129345A876
129346578A
12934657A8
1293465A78
1293465A87
129346758A
12934675A8
12934A5678
12934A5687
12934A5867
12934A5876
12934A8567
12934A8576
152346789A
15234678A9
152346798A
15234679A8
152346978A
15234697A8
1523469A78
1523469A87
152349678A
15234967A8
1523496A78
1523496A87
152349A678
152349A687
152349A867
152349A876
152364789A
15236478A9
152364798A
15236479A8
152364978A
15236497A8
1523649A78
1523649A87
152367489A
15236748A9
152367498A
15236749A8
152367849A
15236784A9
1523678A49
152634789A
15263478A9
152634798A
15263479A8
152634978A
15263497A8
1526349A78
1526349A87
152637489A
15263748A9
152637498A
15263749A8
152637849A
15263784A9
1526378A49
152673489A
15267348A9
152673498A
15267349A8
152673849A
15267384A9
1526738A49
152678349A
15267834A9
1526783A49
152678A349
156234789A
15623478A9
156234798A
15623479A8
156234978A
15623497A8
1562349A78
1562349A87
156237489A
15623748A9
156237498A
15623749A8
156237849A
15623784A9
1562378A49
156273489A
15627348A9
156273498A
15627349A8
156273849A
15627384A9
1562738A49
156278349A
15627834A9
1562783A49
156278A349
156723489A
15672348A9
156723498A
15672349A8
156723849A
15672384A9
1567238A49
156728349A
15672834A9
1567283A49
156728A349
156782349A
15678234A9
1567823A49
156782A349
213456789A
21345678A9
213456798A
21345679A8
213456978A
21345697A8
2134569A78
2134569A87
213459678A
21345967A8
2134596A78
2134596A87
213459A678
213459A687
213459A867
213459A876
213495678A
21349567A8
2134956A78
2134956A87
213495A678
213495A687
213495A867
213495A876
213496578A
21349657A8
2134965A78
2134965A87
213496758A
21349675A8
21349A5678
21349A5687
21349A5867
21349A5876
21349A8567
21349A8576
213546789A
21354678A9
213546798A
21354679A8
213546978A
21354697A8
2135469A78
2135469A87
213549678A
21354967A8
2135496A78
2135496A87
213549A678
213549A687
213549A867
213549A876
213564789A
21356478A9
213564798A
21356479A8
213564978A
21356497A8
2135649A78
2135649A87
213567489A
21356748A9
213567498A
21356749A8
213567849A
21356784A9
2135678A49
213945678A
21394567A8
2139456A78
2139456A87
213945A678
213945A687
213945A867
213945A876
213946578A
21394657A8
2139465A78
2139465A87
213946758A
21394675A8
21394A5678
21394A5687
21394A5867
21394A5876
21394A8567
21394A8576
215346789A
21534678A9
215346798A
21534679A8
215346978A
21534697A8
2153469A78
2153469A87
215349678A
21534967A8
2153496A78
2153496A87
215349A678
215349A687
215349A867
215349A876
215364789A
21536478A9
215364798A
21536479A8
215364978A
21536497A8
2153649A78
2153649A87
215367489A
21536748A9
215367498A
21536749A8
215367849A
21536784A9
2153678A49
215634789A
21563478A9
215634798A
21563479A8
215634978A
21563497A8
2156349A78
2156349A87
215637489A
21563748A9
215637498A
21563749A8
215637849A
21563784A9
2156378A49
215673489A
21567348A9
215673498A
21567349A8
215673849A
21567384A9
2156738A49
215678349A
21567834A9
2156783A49
215678A349
219345678A
21934567A8
2193456A78
2193456A87
219345A678
219345A687
219345A867
219345A876
219346578A
21934657A8
2193465A78
2193465A87
219346758A
21934675A8
21934A5678
21934A5687
21934A5867
21934A5876
21934A8567
21934A8576
231456789A
23145678A9
231456798A
23145679A8
231456978A
23145697A8
2314569A78
2314569A87
231459678A
23145967A8
2314596A78
2314596A87
231459A678
231459A687
231459A867
231459A876
231495678A
23149567A8
2314956A78
2314956A87
231495A678
231495A687
231495A867
231495A876
231496578A
23149657A8
2314965A78
2314965A87
231496758A
23149675A8
23149A5678
23149A5687
23149A5867
23149A5876
23149A8567
23149A8576
231546789A
23154678A9
231546798A
23154679A8
231546978A
23154697A8
2315469A78
2315469A87
231549678A
23154967A8
2315496A78
2315496A87
231549A678
231549A687
231549A867
231549A876
231564789A
23156478A9
231564798A
23156479A8
231564978A
23156497A8
2315649A78
2315649A87
231567489A
23156748A9
231567498A
23156749A8
231567849A
23156784A9
2315678A49
231945678A
23194567A8
2319456A78
2319456A87
231945A678
231945A687
231945A867
231945A876
231946578A
23194657A8
2319465A78
2319465A87
231946758A
23194675A8
23194A5678
23194A5687
23194A5867
23194A5876
23194A8567
23194A8576
234156789A
23415678A9
234156798A
23415679A8
234156978A
23415697A8
2341569A78
2341569A87
234159678A
23415967A8
2341596A78
2341596A87
234159A678
234159A687
234159A867
234159A876
234195678A
23419567A8
2341956A78
2341956A87
234195A678
234195A687
234195A867
234195A876
234196578A
23419657A8
2341965A78
2341965A87
234196758A
23419675A8
23419A5678
23419A5687
23419A5867
23419A5876
23419A8567
23419A8576
234915678A
23491567A8
2349156A78
2349156A87
234915A678
234915A687
234915A867
234915A876
234916578A
23491657A8
2349165A78
2349165A87
234916758A
23491675A8
23491A5678
23491A5687
23491A5867
23491A5876
23491A8567
23491A8576
2349A15678
2349A15687
2349A15867
2349A15876
2349A18567
2349A18576
2349A81567
2349A81576
321456789A
32145678A9
321456798A
32145679A8
321456978A
32145697A8
3214569A78
3214569A87
321459678A
32145967A8
3214596A78
3214596A87
321459A678
321459A687
321459A867
321459A876
321495678A
32149567A8
3214956A78
3214956A87
321495A678
321495A687
321495A867
321495A876
321496578A
32149657A8
3214965A78
3214965A87
321496758A
32149675A8
32149A5678
32149A5687
32149A5867
32149A5876
32149A8567
32149A8576
321546789A
32154678A9
321546798A
32154679A8
321546978A
32154697A8
3215469A78
3215469A87
321549678A
32154967A8
3215496A78
3215496A87
321549A678
321549A687
321549A867
321549A876
321564789A
32156478A9
321564798A
32156479A8
321564978A
32156497A8
3215649A78
3215649A87
321567489A
32156748A9
321567498A
32156749A8
321567849A
32156784A9
3215678A49
321945678A
32194567A8
3219456A78
3219456A87
321945A678
321945A687
321945A867
321945A876
321946578A
32194657A8
3219465A78
3219465A87
321946758A
32194675A8
32194A5678
32194A5687
32194A5867
32194A5876
32194A8567
32194A8576
324156789A
32415678A9
324156798A
32415679A8
324156978A
32415697A8
3241569A78
3241569A87
324159678A
32415967A8
3241596A78
3241596A87
324159A678
324159A687
324159A867
324159A876
324195678A
32419567A8
3241956A78
3241956A87
324195A678
324195A687
324195A867
324195A876
324196578A
32419657A8
3241965A78
3241965A87
324196758A
32419675A8
32419A5678
32419A5687
32419A5867
32419A5876
32419A8567
32419A8576
324915678A
32491567A8
3249156A78
3249156A87
324915A678
324915A687
324915A867
324915A876
324916578A
32491657A8
3249165A78
3249165A87
324916758A
32491675A8
32491A5678
32491A5687
32491A5867
32491A5876
32491A8567
32491A8576
3249A15678
3249A15687
3249A15867
3249A15876
3249A18567
3249A18576
3249A81567
3249A81576
342156789A
34215678A9
342156798A
34215679A8
342156978A
34215697A8
3421569A78
3421569A87
342159678A
34215967A8
3421596A78
3421596A87
342159A678
342159A687
342159A867
342159A876
342195678A
34219567A8
3421956A78
3421956A87
342195A678
342195A687
342195A867
342195A876
342196578A
34219657A8
3421965A78
3421965A87
342196758A
34219675A8
34219A5678
34219A5687
34219A5867
34219A5876
34219A8567
34219A8576
342915678A
34291567A8
3429156A78
3429156A87
342915A678
342915A687
342915A867
342915A876
342916578A
34291657A8
3429165A78
3429165A87
342916758A
34291675A8
34291A5678
34291A5687
34291A5867
34291A5876
34291A8567
34291A8576
3429A15678
3429A15687
3429A15867
3429A15876
3429A18567
3429A18576
3429A81567
3429A81576
349215678A
34921567A8
3492156A78
3492156A87
349215A678
349215A687
349215A867
349215A876
349216578A
34921657A8
3492165A78
3492165A87
349216758A
34921675A8
34921A5678
34921A5687
34921A5867
34921A5876
34921A8567
34921A8576
3492A15678
3492A15687
3492A15867
3492A15876
3492A18567
3492A18576
3492A81567
3492A81576
349A215678
349A215687
349A215867
349A215876
349A218567
349A218576
349A281567
349A281576
432156789A
43215678A9
432156798A
43215679A8
432156978A
43215697A8
4321569A78
4321569A87
432159678A
43215967A8
4321596A78
4321596A87
432159A678
432159A687
432159A867
432159A876
432195678A
43219567A8
4321956A78
4321956A87
432195A678
432195A687
432195A867
432195A876
432196578A
43219657A8
4321965A78
4321965A87
432196758A
43219675A8
43219A5678
43219A5687
43219A5867
43219A5876
43219A8567
43219A8576
432915678A
43291567A8
4329156A78
4329156A87
432915A678
432915A687
432915A867
432915A876
432916578A
43291657A8
4329165A78
4329165A87
432916758A
43291675A8
43291A5678
43291A5687
43291A5867
43291A5876
43291A8567
43291A8576
4329A15678
4329A15687
4329A15867
4329A15876
4329A18567
4329A18576
4329A81567
4329A81576
439215678A
43921567A8
4392156A78
4392156A87
439215A678
439215A687
439215A867
439215A876
439216578A
43921657A8
4392165A78
4392165A87
439216758A
43921675A8
43921A5678
43921A5687
43921A5867
43921A5876
43921A8567
43921A8576
4392A15678
4392A15687
4392A15867
4392A15876
4392A18567
4392A18576
4392A81567
4392A81576
439A215678
439A215687
439A215867
439A215876
439A218567
439A218576
439A281567
439A281576
512346789A
51234678A9
512346798A
51234679A8
512346978A
51234697A8
5123469A78
5123469A87
512349678A
51234967A8
5123496A78
5123496A87
512349A678
512349A687
512349A867
512349A876
512364789A
51236478A9
512364798A
51236479A8
512364978A
51236497A8
5123649A78
5123649A87
512367489A
51236748A9
512367498A
51236749A8
512367849A
51236784A9
5123678A49
512634789A
51263478A9
512634798A
51263479A8
512634978A
51263497A8
5126349A78
5126349A87
512637489A
51263748A9
512637498A
51263749A8
512637849A
51263784A9
5126378A49
512673489A
51267348A9
512673498A
51267349A8
512673849A
51267384A9
5126738A49
512678349A
51267834A9
5126783A49
512678A349
516234789A
51623478A9
516234798A
51623479A8
516234978A
51623497A8
5162349A78
5162349A87
516237489A
51623748A9
516237498A
51623749A8
516237849A
51623784A9
5162378A49
516273489A
51627348A9
516273498A
51627349A8
516273849A
51627384A9
5162738A49
516278349A
51627834A9
5162783A49
516278A349
516723489A
51672348A9
516723498A
51672349A8
516723849A
51672384A9
5167238A49
516728349A
51672834A9
5167283A49
516728A349
516782349A
51678234A9
5167823A49
516782A349
561234789A
56123478A9
561234798A
56123479A8
561234978A
56123497A8
5612349A78
5612349A87
561237489A
56123748A9
561237498A
56123749A8
561237849A
56123784A9
5612378A49
561273489A
56127348A9
561273498A
56127349A8
561273849A
56127384A9
5612738A49
561278349A
56127834A9
5612783A49
561278A349
561723489A
56172348A9
561723498A
56172349A8
561723849A
56172384A9
5617238A49
561728349A
56172834A9
5617283A49
561728A349
561782349A
56178234A9
5617823A49
561782A349
567123489A
56712348A9
567123498A
56712349A8
567123849A
56712384A9
5671238A49
567128349A
56712834A9
5671283A49
567128A349
567182349A
56718234A9
5671823A49
567182A349
\end{spacing}

\subsubsection*{The Condorcet domain of size 2349 on 11 alternatives}
\label{sec:cd_11}

\begin{spacing}{0.8}
\footnotesize
\noindent 
123456789AB
123456789BA
12345678A9B
123456798AB
123456798BA
12345679B8A
12345679BA8
123456978AB
123456978BA
12345697B8A
12345697BA8
1234569B78A
1234569B7A8
1234569BA78
1234569BA87
123459678AB
123459678BA
12345967B8A
12345967BA8
1234596B78A
1234596B7A8
1234596BA78
1234596BA87
123459B678A
123459B67A8
123459B6A78
123459B6A87
123459BA678
123459BA687
123459BA867
123459BA876
123495678AB
123495678BA
12349567B8A
12349567BA8
1234956B78A
1234956B7A8
1234956BA78
1234956BA87
123495B678A
123495B67A8
123495B6A78
123495B6A87
123495BA678
123495BA687
123495BA867
123495BA876
12349B5678A
12349B567A8
12349B56A78
12349B56A87
12349B5A678
12349B5A687
12349B5A867
12349B5A876
12349BA5678
12349BA5687
12349BA5867
12349BA5876
12349BA8567
12349BA8576
12349BA8756
12349BA8765
123945678AB
123945678BA
12394567B8A
12394567BA8
1239456B78A
1239456B7A8
1239456BA78
1239456BA87
123945B678A
123945B67A8
123945B6A78
123945B6A87
123945BA678
123945BA687
123945BA867
123945BA876
12394B5678A
12394B567A8
12394B56A78
12394B56A87
12394B5A678
12394B5A687
12394B5A867
12394B5A876
12394BA5678
12394BA5687
12394BA5867
12394BA5876
12394BA8567
12394BA8576
12394BA8756
12394BA8765
1239B45678A
1239B4567A8
1239B456A78
1239B456A87
1239B45A678
1239B45A687
1239B45A867
1239B45A876
1239B4A5678
1239B4A5687
1239B4A5867
1239B4A5876
1239B4A8567
1239B4A8576
1239B4A8756
1239B4A8765
1239B54678A
1239B5467A8
1239B546A78
1239B546A87
1239B54A678
1239B54A687
1239B54A867
1239B54A876
1239B56478A
1239B5647A8
1239B564A78
1239B564A87
1239BA45678
1239BA45687
1239BA45867
1239BA45876
1239BA48567
1239BA48576
1239BA48756
1239BA48765
1239BA84567
1239BA84576
1239BA84756
1239BA84765
1239BA87456
1239BA87465
123B945678A
123B94567A8
123B9456A78
123B9456A87
123B945A678
123B945A687
123B945A867
123B945A876
123B94A5678
123B94A5687
123B94A5867
123B94A5876
123B94A8567
123B94A8576
123B94A8756
123B94A8765
123B954678A
123B95467A8
123B9546A78
123B9546A87
123B954A678
123B954A687
123B954A867
123B954A876
123B956478A
123B95647A8
123B9564A78
123B9564A87
123B9A45678
123B9A45687
123B9A45867
123B9A45876
123B9A48567
123B9A48576
123B9A48756
123B9A48765
123B9A84567
123B9A84576
123B9A84756
123B9A84765
123B9A87456
123B9A87465
124356789AB
124356789BA
12435678A9B
124356798AB
124356798BA
12435679B8A
12435679BA8
124356978AB
124356978BA
12435697B8A
12435697BA8
1243569B78A
1243569B7A8
1243569BA78
1243569BA87
124359678AB
124359678BA
12435967B8A
12435967BA8
1243596B78A
1243596B7A8
1243596BA78
1243596BA87
124359B678A
124359B67A8
124359B6A78
124359B6A87
124359BA678
124359BA687
124359BA867
124359BA876
124395678AB
124395678BA
12439567B8A
12439567BA8
1243956B78A
1243956B7A8
1243956BA78
1243956BA87
124395B678A
124395B67A8
124395B6A78
124395B6A87
124395BA678
124395BA687
124395BA867
124395BA876
12439B5678A
12439B567A8
12439B56A78
12439B56A87
12439B5A678
12439B5A687
12439B5A867
12439B5A876
12439BA5678
12439BA5687
12439BA5867
12439BA5876
12439BA8567
12439BA8576
12439BA8756
12439BA8765
124536789AB
124536789BA
12453678A9B
124536798AB
124536798BA
12453679B8A
12453679BA8
124536978AB
124536978BA
12453697B8A
12453697BA8
1245369B78A
1245369B7A8
1245369BA78
1245369BA87
124539678AB
124539678BA
12453967B8A
12453967BA8
1245396B78A
1245396B7A8
1245396BA78
1245396BA87
124539B678A
124539B67A8
124539B6A78
124539B6A87
124539BA678
124539BA687
124539BA867
124539BA876
124563789AB
124563789BA
12456378A9B
124563798AB
124563798BA
12456379B8A
12456379BA8
124563978AB
124563978BA
12456397B8A
12456397BA8
1245639B78A
1245639B7A8
1245639BA78
1245639BA87
124567389AB
124567389BA
12456738A9B
124567398AB
124567398BA
12456739B8A
12456739BA8
124567839AB
124567839BA
12456783A9B
1245678A39B
132456789AB
132456789BA
13245678A9B
132456798AB
132456798BA
13245679B8A
13245679BA8
132456978AB
132456978BA
13245697B8A
13245697BA8
1324569B78A
1324569B7A8
1324569BA78
1324569BA87
132459678AB
132459678BA
13245967B8A
13245967BA8
1324596B78A
1324596B7A8
1324596BA78
1324596BA87
132459B678A
132459B67A8
132459B6A78
132459B6A87
132459BA678
132459BA687
132459BA867
132459BA876
132495678AB
132495678BA
13249567B8A
13249567BA8
1324956B78A
1324956B7A8
1324956BA78
1324956BA87
132495B678A
132495B67A8
132495B6A78
132495B6A87
132495BA678
132495BA687
132495BA867
132495BA876
13249B5678A
13249B567A8
13249B56A78
13249B56A87
13249B5A678
13249B5A687
13249B5A867
13249B5A876
13249BA5678
13249BA5687
13249BA5867
13249BA5876
13249BA8567
13249BA8576
13249BA8756
13249BA8765
132945678AB
132945678BA
13294567B8A
13294567BA8
1329456B78A
1329456B7A8
1329456BA78
1329456BA87
132945B678A
132945B67A8
132945B6A78
132945B6A87
132945BA678
132945BA687
132945BA867
132945BA876
13294B5678A
13294B567A8
13294B56A78
13294B56A87
13294B5A678
13294B5A687
13294B5A867
13294B5A876
13294BA5678
13294BA5687
13294BA5867
13294BA5876
13294BA8567
13294BA8576
13294BA8756
13294BA8765
1329B45678A
1329B4567A8
1329B456A78
1329B456A87
1329B45A678
1329B45A687
1329B45A867
1329B45A876
1329B4A5678
1329B4A5687
1329B4A5867
1329B4A5876
1329B4A8567
1329B4A8576
1329B4A8756
1329B4A8765
1329B54678A
1329B5467A8
1329B546A78
1329B546A87
1329B54A678
1329B54A687
1329B54A867
1329B54A876
1329B56478A
1329B5647A8
1329B564A78
1329B564A87
1329BA45678
1329BA45687
1329BA45867
1329BA45876
1329BA48567
1329BA48576
1329BA48756
1329BA48765
1329BA84567
1329BA84576
1329BA84756
1329BA84765
1329BA87456
1329BA87465
132B945678A
132B94567A8
132B9456A78
132B9456A87
132B945A678
132B945A687
132B945A867
132B945A876
132B94A5678
132B94A5687
132B94A5867
132B94A5876
132B94A8567
132B94A8576
132B94A8756
132B94A8765
132B954678A
132B95467A8
132B9546A78
132B9546A87
132B954A678
132B954A687
132B954A867
132B954A876
132B956478A
132B95647A8
132B9564A78
132B9564A87
132B9A45678
132B9A45687
132B9A45867
132B9A45876
132B9A48567
132B9A48576
132B9A48756
132B9A48765
132B9A84567
132B9A84576
132B9A84756
132B9A84765
132B9A87456
132B9A87465
139245678AB
139245678BA
13924567B8A
13924567BA8
1392456B78A
1392456B7A8
1392456BA78
1392456BA87
139245B678A
139245B67A8
139245B6A78
139245B6A87
139245BA678
139245BA687
139245BA867
139245BA876
13924B5678A
13924B567A8
13924B56A78
13924B56A87
13924B5A678
13924B5A687
13924B5A867
13924B5A876
13924BA5678
13924BA5687
13924BA5867
13924BA5876
13924BA8567
13924BA8576
13924BA8756
13924BA8765
1392B45678A
1392B4567A8
1392B456A78
1392B456A87
1392B45A678
1392B45A687
1392B45A867
1392B45A876
1392B4A5678
1392B4A5687
1392B4A5867
1392B4A5876
1392B4A8567
1392B4A8576
1392B4A8756
1392B4A8765
1392B54678A
1392B5467A8
1392B546A78
1392B546A87
1392B54A678
1392B54A687
1392B54A867
1392B54A876
1392B56478A
1392B5647A8
1392B564A78
1392B564A87
1392BA45678
1392BA45687
1392BA45867
1392BA45876
1392BA48567
1392BA48576
1392BA48756
1392BA48765
1392BA84567
1392BA84576
1392BA84756
1392BA84765
1392BA87456
1392BA87465
139B245678A
139B24567A8
139B2456A78
139B2456A87
139B245A678
139B245A687
139B245A867
139B245A876
139B24A5678
139B24A5687
139B24A5867
139B24A5876
139B24A8567
139B24A8576
139B24A8756
139B24A8765
139B254678A
139B25467A8
139B2546A78
139B2546A87
139B254A678
139B254A687
139B254A867
139B254A876
139B256478A
139B25647A8
139B2564A78
139B2564A87
139B2A45678
139B2A45687
139B2A45867
139B2A45876
139B2A48567
139B2A48576
139B2A48756
139B2A48765
139B2A84567
139B2A84576
139B2A84756
139B2A84765
139B2A87456
139B2A87465
139BA245678
139BA245687
139BA245867
139BA245876
139BA248567
139BA248576
139BA248756
139BA248765
139BA284567
139BA284576
139BA284756
139BA284765
139BA287456
139BA287465
139BA824567
139BA824576
139BA824756
139BA824765
139BA827456
139BA827465
139BA872456
139BA872465
213456789AB
213456789BA
21345678A9B
213456798AB
213456798BA
21345679B8A
21345679BA8
213456978AB
213456978BA
21345697B8A
21345697BA8
2134569B78A
2134569B7A8
2134569BA78
2134569BA87
213459678AB
213459678BA
21345967B8A
21345967BA8
2134596B78A
2134596B7A8
2134596BA78
2134596BA87
213459B678A
213459B67A8
213459B6A78
213459B6A87
213459BA678
213459BA687
213459BA867
213459BA876
213495678AB
213495678BA
21349567B8A
21349567BA8
2134956B78A
2134956B7A8
2134956BA78
2134956BA87
213495B678A
213495B67A8
213495B6A78
213495B6A87
213495BA678
213495BA687
213495BA867
213495BA876
21349B5678A
21349B567A8
21349B56A78
21349B56A87
21349B5A678
21349B5A687
21349B5A867
21349B5A876
21349BA5678
21349BA5687
21349BA5867
21349BA5876
21349BA8567
21349BA8576
21349BA8756
21349BA8765
213945678AB
213945678BA
21394567B8A
21394567BA8
2139456B78A
2139456B7A8
2139456BA78
2139456BA87
213945B678A
213945B67A8
213945B6A78
213945B6A87
213945BA678
213945BA687
213945BA867
213945BA876
21394B5678A
21394B567A8
21394B56A78
21394B56A87
21394B5A678
21394B5A687
21394B5A867
21394B5A876
21394BA5678
21394BA5687
21394BA5867
21394BA5876
21394BA8567
21394BA8576
21394BA8756
21394BA8765
2139B45678A
2139B4567A8
2139B456A78
2139B456A87
2139B45A678
2139B45A687
2139B45A867
2139B45A876
2139B4A5678
2139B4A5687
2139B4A5867
2139B4A5876
2139B4A8567
2139B4A8576
2139B4A8756
2139B4A8765
2139B54678A
2139B5467A8
2139B546A78
2139B546A87
2139B54A678
2139B54A687
2139B54A867
2139B54A876
2139B56478A
2139B5647A8
2139B564A78
2139B564A87
2139BA45678
2139BA45687
2139BA45867
2139BA45876
2139BA48567
2139BA48576
2139BA48756
2139BA48765
2139BA84567
2139BA84576
2139BA84756
2139BA84765
2139BA87456
2139BA87465
213B945678A
213B94567A8
213B9456A78
213B9456A87
213B945A678
213B945A687
213B945A867
213B945A876
213B94A5678
213B94A5687
213B94A5867
213B94A5876
213B94A8567
213B94A8576
213B94A8756
213B94A8765
213B954678A
213B95467A8
213B9546A78
213B9546A87
213B954A678
213B954A687
213B954A867
213B954A876
213B956478A
213B95647A8
213B9564A78
213B9564A87
213B9A45678
213B9A45687
213B9A45867
213B9A45876
213B9A48567
213B9A48576
213B9A48756
213B9A48765
213B9A84567
213B9A84576
213B9A84756
213B9A84765
213B9A87456
213B9A87465
214356789AB
214356789BA
21435678A9B
214356798AB
214356798BA
21435679B8A
21435679BA8
214356978AB
214356978BA
21435697B8A
21435697BA8
2143569B78A
2143569B7A8
2143569BA78
2143569BA87
214359678AB
214359678BA
21435967B8A
21435967BA8
2143596B78A
2143596B7A8
2143596BA78
2143596BA87
214359B678A
214359B67A8
214359B6A78
214359B6A87
214359BA678
214359BA687
214359BA867
214359BA876
214395678AB
214395678BA
21439567B8A
21439567BA8
2143956B78A
2143956B7A8
2143956BA78
2143956BA87
214395B678A
214395B67A8
214395B6A78
214395B6A87
214395BA678
214395BA687
214395BA867
214395BA876
21439B5678A
21439B567A8
21439B56A78
21439B56A87
21439B5A678
21439B5A687
21439B5A867
21439B5A876
21439BA5678
21439BA5687
21439BA5867
21439BA5876
21439BA8567
21439BA8576
21439BA8756
21439BA8765
214536789AB
214536789BA
21453678A9B
214536798AB
214536798BA
21453679B8A
21453679BA8
214536978AB
214536978BA
21453697B8A
21453697BA8
2145369B78A
2145369B7A8
2145369BA78
2145369BA87
214539678AB
214539678BA
21453967B8A
21453967BA8
2145396B78A
2145396B7A8
2145396BA78
2145396BA87
214539B678A
214539B67A8
214539B6A78
214539B6A87
214539BA678
214539BA687
214539BA867
214539BA876
214563789AB
214563789BA
21456378A9B
214563798AB
214563798BA
21456379B8A
21456379BA8
214563978AB
214563978BA
21456397B8A
21456397BA8
2145639B78A
2145639B7A8
2145639BA78
2145639BA87
214567389AB
214567389BA
21456738A9B
214567398AB
214567398BA
21456739B8A
21456739BA8
214567839AB
214567839BA
21456783A9B
2145678A39B
241356789AB
241356789BA
24135678A9B
241356798AB
241356798BA
24135679B8A
24135679BA8
241356978AB
241356978BA
24135697B8A
24135697BA8
2413569B78A
2413569B7A8
2413569BA78
2413569BA87
241359678AB
241359678BA
24135967B8A
24135967BA8
2413596B78A
2413596B7A8
2413596BA78
2413596BA87
241359B678A
241359B67A8
241359B6A78
241359B6A87
241359BA678
241359BA687
241359BA867
241359BA876
241395678AB
241395678BA
24139567B8A
24139567BA8
2413956B78A
2413956B7A8
2413956BA78
2413956BA87
241395B678A
241395B67A8
241395B6A78
241395B6A87
241395BA678
241395BA687
241395BA867
241395BA876
24139B5678A
24139B567A8
24139B56A78
24139B56A87
24139B5A678
24139B5A687
24139B5A867
24139B5A876
24139BA5678
24139BA5687
24139BA5867
24139BA5876
24139BA8567
24139BA8576
24139BA8756
24139BA8765
241536789AB
241536789BA
24153678A9B
241536798AB
241536798BA
24153679B8A
24153679BA8
241536978AB
241536978BA
24153697B8A
24153697BA8
2415369B78A
2415369B7A8
2415369BA78
2415369BA87
241539678AB
241539678BA
24153967B8A
24153967BA8
2415396B78A
2415396B7A8
2415396BA78
2415396BA87
241539B678A
241539B67A8
241539B6A78
241539B6A87
241539BA678
241539BA687
241539BA867
241539BA876
241563789AB
241563789BA
24156378A9B
241563798AB
241563798BA
24156379B8A
24156379BA8
241563978AB
241563978BA
24156397B8A
24156397BA8
2415639B78A
2415639B7A8
2415639BA78
2415639BA87
241567389AB
241567389BA
24156738A9B
241567398AB
241567398BA
24156739B8A
24156739BA8
241567839AB
241567839BA
24156783A9B
2415678A39B
245136789AB
245136789BA
24513678A9B
245136798AB
245136798BA
24513679B8A
24513679BA8
245136978AB
245136978BA
24513697B8A
24513697BA8
2451369B78A
2451369B7A8
2451369BA78
2451369BA87
245139678AB
245139678BA
24513967B8A
24513967BA8
2451396B78A
2451396B7A8
2451396BA78
2451396BA87
245139B678A
245139B67A8
245139B6A78
245139B6A87
245139BA678
245139BA687
245139BA867
245139BA876
245163789AB
245163789BA
24516378A9B
245163798AB
245163798BA
24516379B8A
24516379BA8
245163978AB
245163978BA
24516397B8A
24516397BA8
2451639B78A
2451639B7A8
2451639BA78
2451639BA87
245167389AB
245167389BA
24516738A9B
245167398AB
245167398BA
24516739B8A
24516739BA8
245167839AB
245167839BA
24516783A9B
2451678A39B
245613789AB
245613789BA
24561378A9B
245613798AB
245613798BA
24561379B8A
24561379BA8
245613978AB
245613978BA
24561397B8A
24561397BA8
2456139B78A
2456139B7A8
2456139BA78
2456139BA87
245617389AB
245617389BA
24561738A9B
245617398AB
245617398BA
24561739B8A
24561739BA8
245617839AB
245617839BA
24561783A9B
2456178A39B
245671389AB
245671389BA
24567138A9B
245671398AB
245671398BA
24567139B8A
24567139BA8
245671839AB
245671839BA
24567183A9B
2456718A39B
312456789AB
312456789BA
31245678A9B
312456798AB
312456798BA
31245679B8A
31245679BA8
312456978AB
312456978BA
31245697B8A
31245697BA8
3124569B78A
3124569B7A8
3124569BA78
3124569BA87
312459678AB
312459678BA
31245967B8A
31245967BA8
3124596B78A
3124596B7A8
3124596BA78
3124596BA87
312459B678A
312459B67A8
312459B6A78
312459B6A87
312459BA678
312459BA687
312459BA867
312459BA876
312495678AB
312495678BA
31249567B8A
31249567BA8
3124956B78A
3124956B7A8
3124956BA78
3124956BA87
312495B678A
312495B67A8
312495B6A78
312495B6A87
312495BA678
312495BA687
312495BA867
312495BA876
31249B5678A
31249B567A8
31249B56A78
31249B56A87
31249B5A678
31249B5A687
31249B5A867
31249B5A876
31249BA5678
31249BA5687
31249BA5867
31249BA5876
31249BA8567
31249BA8576
31249BA8756
31249BA8765
312945678AB
312945678BA
31294567B8A
31294567BA8
3129456B78A
3129456B7A8
3129456BA78
3129456BA87
312945B678A
312945B67A8
312945B6A78
312945B6A87
312945BA678
312945BA687
312945BA867
312945BA876
31294B5678A
31294B567A8
31294B56A78
31294B56A87
31294B5A678
31294B5A687
31294B5A867
31294B5A876
31294BA5678
31294BA5687
31294BA5867
31294BA5876
31294BA8567
31294BA8576
31294BA8756
31294BA8765
3129B45678A
3129B4567A8
3129B456A78
3129B456A87
3129B45A678
3129B45A687
3129B45A867
3129B45A876
3129B4A5678
3129B4A5687
3129B4A5867
3129B4A5876
3129B4A8567
3129B4A8576
3129B4A8756
3129B4A8765
3129B54678A
3129B5467A8
3129B546A78
3129B546A87
3129B54A678
3129B54A687
3129B54A867
3129B54A876
3129B56478A
3129B5647A8
3129B564A78
3129B564A87
3129BA45678
3129BA45687
3129BA45867
3129BA45876
3129BA48567
3129BA48576
3129BA48756
3129BA48765
3129BA84567
3129BA84576
3129BA84756
3129BA84765
3129BA87456
3129BA87465
312B945678A
312B94567A8
312B9456A78
312B9456A87
312B945A678
312B945A687
312B945A867
312B945A876
312B94A5678
312B94A5687
312B94A5867
312B94A5876
312B94A8567
312B94A8576
312B94A8756
312B94A8765
312B954678A
312B95467A8
312B9546A78
312B9546A87
312B954A678
312B954A687
312B954A867
312B954A876
312B956478A
312B95647A8
312B9564A78
312B9564A87
312B9A45678
312B9A45687
312B9A45867
312B9A45876
312B9A48567
312B9A48576
312B9A48756
312B9A48765
312B9A84567
312B9A84576
312B9A84756
312B9A84765
312B9A87456
312B9A87465
319245678AB
319245678BA
31924567B8A
31924567BA8
3192456B78A
3192456B7A8
3192456BA78
3192456BA87
319245B678A
319245B67A8
319245B6A78
319245B6A87
319245BA678
319245BA687
319245BA867
319245BA876
31924B5678A
31924B567A8
31924B56A78
31924B56A87
31924B5A678
31924B5A687
31924B5A867
31924B5A876
31924BA5678
31924BA5687
31924BA5867
31924BA5876
31924BA8567
31924BA8576
31924BA8756
31924BA8765
3192B45678A
3192B4567A8
3192B456A78
3192B456A87
3192B45A678
3192B45A687
3192B45A867
3192B45A876
3192B4A5678
3192B4A5687
3192B4A5867
3192B4A5876
3192B4A8567
3192B4A8576
3192B4A8756
3192B4A8765
3192B54678A
3192B5467A8
3192B546A78
3192B546A87
3192B54A678
3192B54A687
3192B54A867
3192B54A876
3192B56478A
3192B5647A8
3192B564A78
3192B564A87
3192BA45678
3192BA45687
3192BA45867
3192BA45876
3192BA48567
3192BA48576
3192BA48756
3192BA48765
3192BA84567
3192BA84576
3192BA84756
3192BA84765
3192BA87456
3192BA87465
319B245678A
319B24567A8
319B2456A78
319B2456A87
319B245A678
319B245A687
319B245A867
319B245A876
319B24A5678
319B24A5687
319B24A5867
319B24A5876
319B24A8567
319B24A8576
319B24A8756
319B24A8765
319B254678A
319B25467A8
319B2546A78
319B2546A87
319B254A678
319B254A687
319B254A867
319B254A876
319B256478A
319B25647A8
319B2564A78
319B2564A87
319B2A45678
319B2A45687
319B2A45867
319B2A45876
319B2A48567
319B2A48576
319B2A48756
319B2A48765
319B2A84567
319B2A84576
319B2A84756
319B2A84765
319B2A87456
319B2A87465
319BA245678
319BA245687
319BA245867
319BA245876
319BA248567
319BA248576
319BA248756
319BA248765
319BA284567
319BA284576
319BA284756
319BA284765
319BA287456
319BA287465
319BA824567
319BA824576
319BA824756
319BA824765
319BA827456
319BA827465
319BA872456
319BA872465
391245678AB
391245678BA
39124567B8A
39124567BA8
3912456B78A
3912456B7A8
3912456BA78
3912456BA87
391245B678A
391245B67A8
391245B6A78
391245B6A87
391245BA678
391245BA687
391245BA867
391245BA876
39124B5678A
39124B567A8
39124B56A78
39124B56A87
39124B5A678
39124B5A687
39124B5A867
39124B5A876
39124BA5678
39124BA5687
39124BA5867
39124BA5876
39124BA8567
39124BA8576
39124BA8756
39124BA8765
3912B45678A
3912B4567A8
3912B456A78
3912B456A87
3912B45A678
3912B45A687
3912B45A867
3912B45A876
3912B4A5678
3912B4A5687
3912B4A5867
3912B4A5876
3912B4A8567
3912B4A8576
3912B4A8756
3912B4A8765
3912B54678A
3912B5467A8
3912B546A78
3912B546A87
3912B54A678
3912B54A687
3912B54A867
3912B54A876
3912B56478A
3912B5647A8
3912B564A78
3912B564A87
3912BA45678
3912BA45687
3912BA45867
3912BA45876
3912BA48567
3912BA48576
3912BA48756
3912BA48765
3912BA84567
3912BA84576
3912BA84756
3912BA84765
3912BA87456
3912BA87465
391B245678A
391B24567A8
391B2456A78
391B2456A87
391B245A678
391B245A687
391B245A867
391B245A876
391B24A5678
391B24A5687
391B24A5867
391B24A5876
391B24A8567
391B24A8576
391B24A8756
391B24A8765
391B254678A
391B25467A8
391B2546A78
391B2546A87
391B254A678
391B254A687
391B254A867
391B254A876
391B256478A
391B25647A8
391B2564A78
391B2564A87
391B2A45678
391B2A45687
391B2A45867
391B2A45876
391B2A48567
391B2A48576
391B2A48756
391B2A48765
391B2A84567
391B2A84576
391B2A84756
391B2A84765
391B2A87456
391B2A87465
391BA245678
391BA245687
391BA245867
391BA245876
391BA248567
391BA248576
391BA248756
391BA248765
391BA284567
391BA284576
391BA284756
391BA284765
391BA287456
391BA287465
391BA824567
391BA824576
391BA824756
391BA824765
391BA827456
391BA827465
391BA872456
391BA872465
39B1245678A
39B124567A8
39B12456A78
39B12456A87
39B1245A678
39B1245A687
39B1245A867
39B1245A876
39B124A5678
39B124A5687
39B124A5867
39B124A5876
39B124A8567
39B124A8576
39B124A8756
39B124A8765
39B1254678A
39B125467A8
39B12546A78
39B12546A87
39B1254A678
39B1254A687
39B1254A867
39B1254A876
39B1256478A
39B125647A8
39B12564A78
39B12564A87
39B12A45678
39B12A45687
39B12A45867
39B12A45876
39B12A48567
39B12A48576
39B12A48756
39B12A48765
39B12A84567
39B12A84576
39B12A84756
39B12A84765
39B12A87456
39B12A87465
39B1A245678
39B1A245687
39B1A245867
39B1A245876
39B1A248567
39B1A248576
39B1A248756
39B1A248765
39B1A284567
39B1A284576
39B1A284756
39B1A284765
39B1A287456
39B1A287465
39B1A824567
39B1A824576
39B1A824756
39B1A824765
39B1A827456
39B1A827465
39B1A872456
39B1A872465
39BA1245678
39BA1245687
39BA1245867
39BA1245876
39BA1248567
39BA1248576
39BA1248756
39BA1248765
39BA1284567
39BA1284576
39BA1284756
39BA1284765
39BA1287456
39BA1287465
39BA1824567
39BA1824576
39BA1824756
39BA1824765
39BA1827456
39BA1827465
39BA1872456
39BA1872465
39BA8124567
39BA8124576
39BA8124756
39BA8124765
39BA8127456
39BA8127465
39BA8172456
39BA8172465
421356789AB
421356789BA
42135678A9B
421356798AB
421356798BA
42135679B8A
42135679BA8
421356978AB
421356978BA
42135697B8A
42135697BA8
4213569B78A
4213569B7A8
4213569BA78
4213569BA87
421359678AB
421359678BA
42135967B8A
42135967BA8
4213596B78A
4213596B7A8
4213596BA78
4213596BA87
421359B678A
421359B67A8
421359B6A78
421359B6A87
421359BA678
421359BA687
421359BA867
421359BA876
421395678AB
421395678BA
42139567B8A
42139567BA8
4213956B78A
4213956B7A8
4213956BA78
4213956BA87
421395B678A
421395B67A8
421395B6A78
421395B6A87
421395BA678
421395BA687
421395BA867
421395BA876
42139B5678A
42139B567A8
42139B56A78
42139B56A87
42139B5A678
42139B5A687
42139B5A867
42139B5A876
42139BA5678
42139BA5687
42139BA5867
42139BA5876
42139BA8567
42139BA8576
42139BA8756
42139BA8765
421536789AB
421536789BA
42153678A9B
421536798AB
421536798BA
42153679B8A
42153679BA8
421536978AB
421536978BA
42153697B8A
42153697BA8
4215369B78A
4215369B7A8
4215369BA78
4215369BA87
421539678AB
421539678BA
42153967B8A
42153967BA8
4215396B78A
4215396B7A8
4215396BA78
4215396BA87
421539B678A
421539B67A8
421539B6A78
421539B6A87
421539BA678
421539BA687
421539BA867
421539BA876
421563789AB
421563789BA
42156378A9B
421563798AB
421563798BA
42156379B8A
42156379BA8
421563978AB
421563978BA
42156397B8A
42156397BA8
4215639B78A
4215639B7A8
4215639BA78
4215639BA87
421567389AB
421567389BA
42156738A9B
421567398AB
421567398BA
42156739B8A
42156739BA8
421567839AB
421567839BA
42156783A9B
4215678A39B
425136789AB
425136789BA
42513678A9B
425136798AB
425136798BA
42513679B8A
42513679BA8
425136978AB
425136978BA
42513697B8A
42513697BA8
4251369B78A
4251369B7A8
4251369BA78
4251369BA87
425139678AB
425139678BA
42513967B8A
42513967BA8
4251396B78A
4251396B7A8
4251396BA78
4251396BA87
425139B678A
425139B67A8
425139B6A78
425139B6A87
425139BA678
425139BA687
425139BA867
425139BA876
425163789AB
425163789BA
42516378A9B
425163798AB
425163798BA
42516379B8A
42516379BA8
425163978AB
425163978BA
42516397B8A
42516397BA8
4251639B78A
4251639B7A8
4251639BA78
4251639BA87
425167389AB
425167389BA
42516738A9B
425167398AB
425167398BA
42516739B8A
42516739BA8
425167839AB
425167839BA
42516783A9B
4251678A39B
425613789AB
425613789BA
42561378A9B
425613798AB
425613798BA
42561379B8A
42561379BA8
425613978AB
425613978BA
42561397B8A
42561397BA8
4256139B78A
4256139B7A8
4256139BA78
4256139BA87
425617389AB
425617389BA
42561738A9B
425617398AB
425617398BA
42561739B8A
42561739BA8
425617839AB
425617839BA
42561783A9B
4256178A39B
425671389AB
425671389BA
42567138A9B
425671398AB
425671398BA
42567139B8A
42567139BA8
425671839AB
425671839BA
42567183A9B
4256718A39B
452136789AB
452136789BA
45213678A9B
452136798AB
452136798BA
45213679B8A
45213679BA8
452136978AB
452136978BA
45213697B8A
45213697BA8
4521369B78A
4521369B7A8
4521369BA78
4521369BA87
452139678AB
452139678BA
45213967B8A
45213967BA8
4521396B78A
4521396B7A8
4521396BA78
4521396BA87
452139B678A
452139B67A8
452139B6A78
452139B6A87
452139BA678
452139BA687
452139BA867
452139BA876
452163789AB
452163789BA
45216378A9B
452163798AB
452163798BA
45216379B8A
45216379BA8
452163978AB
452163978BA
45216397B8A
45216397BA8
4521639B78A
4521639B7A8
4521639BA78
4521639BA87
452167389AB
452167389BA
45216738A9B
452167398AB
452167398BA
45216739B8A
45216739BA8
452167839AB
452167839BA
45216783A9B
4521678A39B
452613789AB
452613789BA
45261378A9B
452613798AB
452613798BA
45261379B8A
45261379BA8
452613978AB
452613978BA
45261397B8A
45261397BA8
4526139B78A
4526139B7A8
4526139BA78
4526139BA87
452617389AB
452617389BA
45261738A9B
452617398AB
452617398BA
45261739B8A
45261739BA8
452617839AB
452617839BA
45261783A9B
4526178A39B
452671389AB
452671389BA
45267138A9B
452671398AB
452671398BA
45267139B8A
45267139BA8
452671839AB
452671839BA
45267183A9B
4526718A39B
456213789AB
456213789BA
45621378A9B
456213798AB
456213798BA
45621379B8A
45621379BA8
456213978AB
456213978BA
45621397B8A
45621397BA8
4562139B78A
4562139B7A8
4562139BA78
4562139BA87
456217389AB
456217389BA
45621738A9B
456217398AB
456217398BA
45621739B8A
45621739BA8
456217839AB
456217839BA
45621783A9B
4562178A39B
456271389AB
456271389BA
45627138A9B
456271398AB
456271398BA
45627139B8A
45627139BA8
456271839AB
456271839BA
45627183A9B
4562718A39B
931245678AB
931245678BA
93124567B8A
93124567BA8
9312456B78A
9312456B7A8
9312456BA78
9312456BA87
931245B678A
931245B67A8
931245B6A78
931245B6A87
931245BA678
931245BA687
931245BA867
931245BA876
93124B5678A
93124B567A8
93124B56A78
93124B56A87
93124B5A678
93124B5A687
93124B5A867
93124B5A876
93124BA5678
93124BA5687
93124BA5867
93124BA5876
93124BA8567
93124BA8576
93124BA8756
93124BA8765
9312B45678A
9312B4567A8
9312B456A78
9312B456A87
9312B45A678
9312B45A687
9312B45A867
9312B45A876
9312B4A5678
9312B4A5687
9312B4A5867
9312B4A5876
9312B4A8567
9312B4A8576
9312B4A8756
9312B4A8765
9312B54678A
9312B5467A8
9312B546A78
9312B546A87
9312B54A678
9312B54A687
9312B54A867
9312B54A876
9312B56478A
9312B5647A8
9312B564A78
9312B564A87
9312BA45678
9312BA45687
9312BA45867
9312BA45876
9312BA48567
9312BA48576
9312BA48756
9312BA48765
9312BA84567
9312BA84576
9312BA84756
9312BA84765
9312BA87456
9312BA87465
931B245678A
931B24567A8
931B2456A78
931B2456A87
931B245A678
931B245A687
931B245A867
931B245A876
931B24A5678
931B24A5687
931B24A5867
931B24A5876
931B24A8567
931B24A8576
931B24A8756
931B24A8765
931B254678A
931B25467A8
931B2546A78
931B2546A87
931B254A678
931B254A687
931B254A867
931B254A876
931B256478A
931B25647A8
931B2564A78
931B2564A87
931B2A45678
931B2A45687
931B2A45867
931B2A45876
931B2A48567
931B2A48576
931B2A48756
931B2A48765
931B2A84567
931B2A84576
931B2A84756
931B2A84765
931B2A87456
931B2A87465
931BA245678
931BA245687
931BA245867
931BA245876
931BA248567
931BA248576
931BA248756
931BA248765
931BA284567
931BA284576
931BA284756
931BA284765
931BA287456
931BA287465
931BA824567
931BA824576
931BA824756
931BA824765
931BA827456
931BA827465
931BA872456
931BA872465
93B1245678A
93B124567A8
93B12456A78
93B12456A87
93B1245A678
93B1245A687
93B1245A867
93B1245A876
93B124A5678
93B124A5687
93B124A5867
93B124A5876
93B124A8567
93B124A8576
93B124A8756
93B124A8765
93B1254678A
93B125467A8
93B12546A78
93B12546A87
93B1254A678
93B1254A687
93B1254A867
93B1254A876
93B1256478A
93B125647A8
93B12564A78
93B12564A87
93B12A45678
93B12A45687
93B12A45867
93B12A45876
93B12A48567
93B12A48576
93B12A48756
93B12A48765
93B12A84567
93B12A84576
93B12A84756
93B12A84765
93B12A87456
93B12A87465
93B1A245678
93B1A245687
93B1A245867
93B1A245876
93B1A248567
93B1A248576
93B1A248756
93B1A248765
93B1A284567
93B1A284576
93B1A284756
93B1A284765
93B1A287456
93B1A287465
93B1A824567
93B1A824576
93B1A824756
93B1A824765
93B1A827456
93B1A827465
93B1A872456
93B1A872465
93BA1245678
93BA1245687
93BA1245867
93BA1245876
93BA1248567
93BA1248576
93BA1248756
93BA1248765
93BA1284567
93BA1284576
93BA1284756
93BA1284765
93BA1287456
93BA1287465
93BA1824567
93BA1824576
93BA1824756
93BA1824765
93BA1827456
93BA1827465
93BA1872456
93BA1872465
93BA8124567
93BA8124576
93BA8124756
93BA8124765
93BA8127456
93BA8127465
93BA8172456
93BA8172465
9B31245678A
9B3124567A8
9B312456A78
9B312456A87
9B31245A678
9B31245A687
9B31245A867
9B31245A876
9B3124A5678
9B3124A5687
9B3124A5867
9B3124A5876
9B3124A8567
9B3124A8576
9B3124A8756
9B3124A8765
9B31254678A
9B3125467A8
9B312546A78
9B312546A87
9B31254A678
9B31254A687
9B31254A867
9B31254A876
9B31256478A
9B3125647A8
9B312564A78
9B312564A87
9B312A45678
9B312A45687
9B312A45867
9B312A45876
9B312A48567
9B312A48576
9B312A48756
9B312A48765
9B312A84567
9B312A84576
9B312A84756
9B312A84765
9B312A87456
9B312A87465
9B31A245678
9B31A245687
9B31A245867
9B31A245876
9B31A248567
9B31A248576
9B31A248756
9B31A248765
9B31A284567
9B31A284576
9B31A284756
9B31A284765
9B31A287456
9B31A287465
9B31A824567
9B31A824576
9B31A824756
9B31A824765
9B31A827456
9B31A827465
9B31A872456
9B31A872465
9B3A1245678
9B3A1245687
9B3A1245867
9B3A1245876
9B3A1248567
9B3A1248576
9B3A1248756
9B3A1248765
9B3A1284567
9B3A1284576
9B3A1284756
9B3A1284765
9B3A1287456
9B3A1287465
9B3A1824567
9B3A1824576
9B3A1824756
9B3A1824765
9B3A1827456
9B3A1827465
9B3A1872456
9B3A1872465
9B3A8124567
9B3A8124576
9B3A8124756
9B3A8124765
9B3A8127456
9B3A8127465
9B3A8172456
9B3A8172465
\end{spacing}

\newpage

\end{document}